\documentclass[conference]{IEEEtran}
\ifCLASSINFOpdf
\else
\fi

\usepackage{epsfig,endnotes}
\usepackage{epsfig,endnotes}
\usepackage{amsfonts}
\usepackage{capt-of}
\usepackage{wrapfig}
\usepackage{wrapfig,booktabs}
\usepackage{lipsum}
\usepackage{adjustbox}
\usepackage{epsfig}
\usepackage{caption}
\usepackage{hhline}
\usepackage{amsmath}
\usepackage{xfrac}
\usepackage{graphicx}
\usepackage[font=footnotesize,caption=false]{subfig}

\usepackage{stmaryrd}
\usepackage{array}
\usepackage{mdwmath}
\usepackage{mdwtab}
\usepackage{eqparbox}
\usepackage{slashbox}
\usepackage{stfloats}
\usepackage{moreverb}
\usepackage{listings}
\usepackage{algpseudocode}
\usepackage{algorithm}
\usepackage{url}
\usepackage{color}
\usepackage{multirow}
\usepackage{setspace}
\usepackage[utf8]{inputenc}
\usepackage[T1]{fontenc}
\usepackage{microtype}
\usepackage{datetime}
\usepackage{epsfig}
\usepackage{wrapfig,lipsum,booktabs}
\usepackage{longtable}
\usepackage{xspace}
\usepackage{tabu}
\usepackage{boldline}
\usepackage{balance}
\usepackage{verbatim}
\usepackage{xcolor}

\usepackage[hidelinks]{hyperref}

\newcommand{\aaf}{\vspace*{-6pt}}
\newcommand{\af}{\vspace*{-3pt}}
\newcommand{\tool}{\textsc{InnerEye}\xspace}
\newcommand{\bbtool}{\textsc{InnerEye-BB}\xspace}
\newcommand{\cctool}{\textsc{InnerEye-CC}\xspace}

\newcommand{\todo}[1]{{\textcolor{black}{#1}}}

\newcommand{\zedit}[1]{{\textcolor{black}{#1}}} 

\newcommand{\ledit}[1]{{\textcolor{black}{#1}}} 

\newtheorem{definition}{Definition}

\begin{document}
\bstctlcite{IEEEexample:BSTcontrol}
%
\title{Neural Machine Translation Inspired Binary Code Similarity Comparison \emph{beyond Function Pairs}}


\author{\IEEEauthorblockN{Fei Zuo\IEEEauthorrefmark{2}, Xiaopeng Li\IEEEauthorrefmark{2}, Patrick Young\IEEEauthorrefmark{3}, Lannan Luo\IEEEauthorrefmark{2}*, Qiang Zeng\IEEEauthorrefmark{2}*, Zhexin Zhang\IEEEauthorrefmark{2}}
               \IEEEauthorblockA{\IEEEauthorrefmark{2}University of South Carolina, \IEEEauthorrefmark{3}Temple University\\
               \{fzuo, xl4, zhexin\}@email.sc.edu, \{lluo, zeng1\}@cse.sc.edu}}

%


\IEEEoverridecommandlockouts
\makeatletter\def\@IEEEpubidpullup{6.5\baselineskip}\makeatother
\IEEEpubid{\parbox{\columnwidth}{
    * Corresponding authors.\\
    Network and Distributed Systems Security (NDSS) Symposium 2019\\    24-27 February 2019, San Diego, CA, USA\\    ISBN 1-891562-55-X\\    https://dx.doi.org/10.14722/ndss.2019.23492\\    www.ndss-symposium.org
}
\hspace{\columnsep}\makebox[\columnwidth]{}}

\maketitle

\begin{abstract}

Binary code analysis allows analyzing binary code without 
having access to the corresponding source code. 
A binary, after disassembly, is expressed in an \emph{assembly language}. This
inspires us to approach binary analysis by leveraging ideas and techniques from
\emph{Natural Language Processing} (NLP), a fruitful area focused on processing text 
of various natural languages. We notice that binary code analysis and NLP share 
many analogical topics, such as semantics extraction, classification, and code/text
comparison. This work \zedit{thus borrows} ideas \zedit{from NLP} to 
address two important code 
similarity comparison problems. (I) Given a pair of basic blocks \zedit{of} different 
instruction set architectures (ISAs), determining whether their semantics \zedit{is} similar; 
and (II) given a piece of code of interest, determining if it is \zedit{\emph{contained}} in 
another piece of code of a different ISA. The solutions to these 
two problems have many applications, such as cross-architecture
vulnerability discovery and code plagiarism detection. 

Despite the evident importance of Problem I, existing solutions are 
either inefficient or imprecise. Inspired by Neural Machine Translation (NMT), which 
is a new approach that tackles text across natural languages very well, we regard
\emph{instructions as words and basic blocks as sentences}, and propose a novel
\emph{cross-(assembly)-lingual} deep learning approach to solving \zedit{Problem I}, 
attaining high efficiency and precision. 
\zedit{Many solutions have been proposed to determine whether two pieces of code,
e.g., functions, are equivalent (called the \emph{equivalence} problem), which is different
from Problem II (called the \emph{containment} problem). 
Resolving the cross-architecture code 
containment problem is a new and more challenging endeavor.
Employing our technique for cross-architecture basic-block comparison, we propose the first 
solution to Problem II.} We implement a prototype system \tool and perform 
a comprehensive evaluation. A comparison between our approach and existing approaches to Problem I shows 
that our system outperforms them in terms of accuracy, efficiency and scalability. 
The case studies \zedit{applying} the system demonstrate that our solution to Problem II is effective.
Moreover, this research showcases how to apply ideas and techniques from NLP to large-scale 
binary code analysis.

\end{abstract}

\section{Introduction} \label{into}

Binary code analysis allows one to analyze binary code without access to the 
corresponding source code. It is widely used for vulnerability discovery, 
code clone detection, user-side crash analysis, etc. 
Today, binary code analysis has become more important than ever. 
Gartner forecasts that 8.4 
billion IoT devices will be in use worldwide in 2017, up 31 percent from 2016, and 
will reach 20.4 billion by 2020~\cite{GartnerIoT}. Due to code reuse and sharing, 
a single vulnerability at source code level may spread across hundreds or more 
devices that have diverse hardware architectures and software platforms~\cite{pewny2015cross}.
However, it is difficult, often unlikely, to obtain the source code from the 
many IoT device companies. \zedit{Thus}, binary code analysis becomes the only feasible approach.

\zedit{Given a code component that is known to contain some vulnerability and a large number
of programs that are compiled for \emph{different} ISAs}, 
by finding programs that contain
similar code components, more instances of the vulnerability can be found. 
\zedit{Thus, cross-architecture binary code analysis draws great interests~\cite{pewny2015cross,eschweiler2016discovre,feng2016scalable,ccs2017graphembedding}.}

\noindent \textbf{Our insight.} 
A binary, after disassembly, is represented in some \emph{assembly language}. This inspires 
us to approach binary code analysis by learning from Natural Language Processing (NLP), a 
fruitful area focused on processing natural language corpora effectively and efficiently. 
Interestingly, the two seemingly remote areas---binary code analysis and NLP---actually share
plenty of analogical topics, such as \emph{semantics extraction from code/text, summarization
of paragraphs/functions, classification of code/articles}, and \emph{code/text similarity comparison}.
We thus propose to \zedit{adapt} the ideas, methods, and techniques used in NLP to resolving binary 
code analysis problems. As \zedit{a showcase}, we \zedit{use this} idea to perform
\emph{cross-architecture binary code similarity comparison}.

\zedit{
Previous work~\cite{pewny2015cross,eschweiler2016discovre, feng2016scalable, ccs2017graphembedding}
essentially resolves the \emph{code equivalence} problem at the function level:
given a pair of functions, it is to determine whether they are equivalent.
We consider a different problem: given a code component, which can be part of a function 
(e.g., the code in a web server that parses the URL) or a set of functions 
(e.g., an implementation of a crypto algorithm), to determine whether
it is \emph{contained} in a program. Thus, it is a \emph{code containment problem}.
The problem has been emphasized by previous work~\cite{jhi2011value,luo2014semantics,ming2016deviation,zhang2014program,wang2009behavior,wang2009detecting,luo2017semantics},
but the proposed solutions can only work for code of the same ISA. Resolving the  cross-architecture  
code  containment  problem  is  a new and important endeavor.  
A solution to this problem is critical for tasks such as fine-grained code plagiarism detection, 
thorough vulnerability search, and virus detection. For example, a code plagiarist
may steal part of a function or a bunch of functions, and 
insert the stolen code into other code; that is, the stolen code is not necessarily a function.
Code plagiarism detection based on searching for equivalent functions is too limited to handle such cases.
}

We define two concrete research problems: (I) given a pair of binary basic blocks \zedit{of} different 
instruction set architectures (ISAs), determining whether their semantics is similar or not; and 
(II) given a piece of critical code, determining \zedit{whether} it is contained in another piece of code 
of a different ISA. Problem I is a \emph{core} \zedit{sub-task} in recent
work on cross-architecture similarity comparison~\cite{pewny2015cross,eschweiler2016discovre, feng2016scalable, ccs2017graphembedding}, while Problem II is new.


\textbf{{\emph{Solution to Problem I.}}}
Problem \textbf{I} is one of the most fundamental problems for code comparison; therefore, many 
approaches have been proposed to resolve it, such as 
fuzzing~\cite{pewny2015cross} and representing a basic block using some features~\cite{eschweiler2016discovre,feng2016scalable, ccs2017graphembedding}.
However, none of existing approaches are able to achieve both high efficiency and precision
for this important problem. 
Fuzzing takes much time to try different inputs, while the feature-based representation is imprecise
(A SVM classifier based {\color{black}on} such features can only achieve AUC = 85\% according to our evaluation).
Given a pair of blocks of different architectures, they, after being disassembled, are
two sequences of instructions in two different assembly languages. 
In light of this insight, we propose to learn from the ideas 
and techniques of Neural Machine Translation (NMT), a new machine translation framework 
based on neural networks proposed by NLP researchers that handles text 
\emph{across} languages very well~\cite{kalchbrenner2013recurrent, sutskever2014sequence}.
NMT frequently uses 
\zedit{word embedding and Long Short-Term Memory (LSTM),
which are capable of \emph{learning features of words and dependencies between words in a sentence and encoding the sentence into a vector representation that captures its semantic 
meaning}}~\cite{palangi2016deep,mueller2016siamese,wieting2015towards}. 
In addition to translating sentences, the NMT model has also been {\color{black}extended} to measure the similarity 
of sentences by comparing their vector representations~\cite{mueller2016siamese,palangi2016deep}.

We regard \emph{instructions as words} and \emph{basic blocks as sentences}, 
and consider that the task of \emph{detecting whether two basic blocks of different ISAs are semantically 
similar} is analogous to \emph{that of determining whether two sentences of different human languages 
have similar meanings}. Following this idea and learning from NMT, we propose a novel neural network-based 
\emph{cross-(assembly)-lingual basic-block embedding model}, which converts a basic block into an \emph{embedding}, 
i.e., a high dimensional numerical vector. The embeddings not only encode basic-block semantics, but 
also capture semantic relationships \emph{across} architectures, such that the similarity of two basic blocks 
can be detected efficiently by measuring the distance between their embeddings.


Recent work~\cite{eschweiler2016discovre,feng2016scalable, ccs2017graphembedding} 
uses several \emph{manually} selected features 
(such as the number of instructions and the number of constants) of a basic block to 
represent it. This inevitably causes 
significant \zedit{information loss in terms of} \emph{the
contained instructions and the dependencies between these instructions}. 
In contrast to using manually selected features, our NMT-inspired approach 
applies deep learning to \emph{automatically} 
capturing such information into a vector. 
Specifically, we propose to employ {\color{black}LSTM}
to automatically 
encode a basic block into an embedding that captures the semantic 
information of the instruction sequence, just like {\color{black}LSTM} is used in NMT to
capture the semantic information of sentences. 
This way, our cross-(assembly)-lingual deep learning approach to Problem \textbf{I} 
achieves both high efficiency and precision \zedit{(AUC = 98\%)}.

{\color{black}\texttt{Gemini}}~\cite{ccs2017graphembedding} also applies neural networks. 
There are two main differences between {\color{black}\texttt{Gemini}} and our work. 
First, \zedit{as described above,} {\color{black}\texttt{Gemini}} uses \emph{manually} selected features to represent a basic block.
Second, {\color{black}\texttt{Gemini}} \zedit{handles the code \emph{equivalence} problem rather than
the code \emph{containment} problem.}

\textbf{{\emph{Solution to Problem II.}}}
\zedit{A special case of Problem \textbf{II}, under the context of 
a single architecture, is well studied~\cite{kamiya2002ccfinder,jiang2007deckard,baker1995finding, schleimer2003winnowing,prechelt2002finding,gao2008binhunt,
luo2014semantics,tamada2004dynamic,schuler2007dynamic,wang2009behavior,jhi2011value}. 
No prior solutions to Problem \textbf{II} under the cross-architecture context exist.
To resolve it,} we decompose the control flow graph (CFG) of the code of interest into 
multiple paths, each of which can be regarded as \emph{a sequence of basic blocks}. Our idea is to 
leverage our solution to Problem \textbf{I} (for efficient and precise basic-block comparison), when 
applying the \emph{Longest Common Subsequence} (LCS) algorithm to comparing the similarity of those 
paths (i.e., basic-block sequences). From there, we can calculate the similarity 
of two pieces of code quantitatively.

\zedit{Note that we do \emph{not} consider an arbitrary piece of code (unless it is a basic block) 
as a sentence, because it should not be 
simply  treated as a straight-line sequence. For example, when a function is invoked, its code is 
not executed sequentially; instead, only a part of the code belonging to a particular \emph{path} gets 
executed. The paths of a function can be scrambled (by compilers) without changing the semantics of the function.} 

We have implemented a prototype system \tool consisting of two \zedit{sub-systems: \bbtool
to resolving Problem \textbf{I} and \cctool to resolving Problem \textbf{II}}. 
We have evaluated \bbtool in terms of accuracy, 
efficiency and scalability, and the evaluation results show that it outperforms existing 
approaches by large margins. Our case studies \zedit{applying} \cctool demonstrate that it can 
successfully resolve cross-architecture code similarity comparison tasks and is much more 
capable than recent work that is limited to comparison of function pairs. 
The datasets, neural network models, and evaluation results are publicly available.\footnote{\url{https://nmt4binaries.github.io}}

We summarize our contributions as follows:

\begin{itemize}

\item
We propose to learn from the successful NMT field to solve the cross-architecture binary 
code similarity \zedit{comparison} problem. \emph{We regard instructions as words and basic blocks 
as sentences.} Thus, the ideas and methodologies for comparing the meanings of sentences 
in different natural languages can be adapted to cross-architecture code similarity comparison.

\item 
We design a precise and efficient cross-(assembly)-lingual basic block embedding model. 
It utilizes \zedit{word embedding} and {\color{black}LSTM}, which are frequently used in NMT, to automatically  
capture \zedit{the semantics and dependencies of instructions}. This is in contrast to prior work 
which \zedit{largely} ignores such information. 

\item
\zedit{We propose the \emph{first} solution to  the cross-architecture code containment
problem. It has many 
applications, such as code plagiarism detection and virus detection.} 

\item 
We implement a prototype \tool and evaluate its accuracy, efficiency, and scalability. 
We use real-world software across architectures to demonstrate the applications 
of \tool.

\item This research successfully demonstrates that it is promising to 
approach binary analysis from the angle of language processing by adapting methodologies, 
ideas and techniques in NLP. 

\end{itemize}


\section{\todo{Related Work}}


\subsection{Traditional Code Similarity \zedit{Comparison}}

\noindent \textbf{Mono-architecture \zedit{solutions}.}
Static plagiarism detection or clone detection includes 
string-based~\cite{baker1995finding,bilenko2003adaptive,deerwester1990indexing}, 
token-based~\cite{kamiya2002ccfinder,schleimer2003winnowing, prechelt2002finding}, 
{\color{black}tree-based}~\cite{jiang2007deckard,koschke2006clone,pewny2014leveraging},
and PDG (program dependence graph)-based~\cite{gabel2008scalable, liu2006gplag,crussell2012attack,li2012cbcd}.
Some approaches are \emph{source code based}, and are less applicable in practice, 
especially concerning closed-source software; e.g., \texttt{CCFINDER}~\cite{kamiya2002ccfinder} 
finds equal suffix chains of source code tokens. 
{\color{black}\texttt{TEDEM}~\cite{pewny2014leveraging} introduces tree edit distances to measure code similarity at the level of basic blocks, which is costly for matching and does not handle all syntactical variation}. Others compare the semantics of binary code using symbolic execution and 
theorem prover, such as \texttt{BinHunt}~\cite{gao2008binhunt} and \texttt{CoP}~\cite{luo2014semantics}, 
but {\color{black}they are computation expensive} 
and thus not applicable for large codebases.

Second, dynamic birthmark based approaches include API birthmark~\cite{tamada2004dynamic,
schuler2007dynamic,chae2013software}
system call birthmark~\cite{wang2009behavior}, function call birthmark~\cite{tamada2004dynamic}, 
instruction birthmark~\cite{tian2013dkisb,park2008detecting}, and core-value birthmark~\cite{jhi2011value}. 
Tamada et al.\ propose an API birthmark for Windows application~\cite{tamada2004dynamic}. 
Schuler et al.\ propose a dynamic birthmark for Java~\cite{schuler2007dynamic}. Wang et 
al.\ introduce two system call based birthmarks suitable for programs invoking 
sufficient system calls~\cite{wang2009behavior}. Jhi et al.\ propose a core-value based 
birthmark for detecting plagiarism~\cite{jhi2011value}. 
However, as they rely on dynamic analysis, \emph{extending them to other architectures and 
adapting to embedded devices would be hard and tedious}.
\zedit{Code coverage of dynamic analysis is another inherent challenge.} 

\noindent \textbf{Cross-architecture \zedit{solutions}.}
Recently, researchers \zedit{start} to address the problem of cross-architecture binary code 
similarity detection. Multi-MH and Multi-k-MH~\cite{pewny2015cross} are the first two methods 
for comparing \emph{function} code across different architectures. However, their fuzzing-based 
basic block similarity comparison and graph (i.e., CFG) matching-based algorithm are too expensive 
to handle \zedit{a large number} of function pairs. 
discovRE~\cite{eschweiler2016discovre} utilizes pre-filtering 
to boost CFG based matching process, but it is still expensive, and the pre-filtering is 
unreliable, outputting too many false negatives. 
{\color{black}Both \texttt{Esh}~\cite{david2016statistical} and its successor~\cite{david2017similarity}
define \emph{strands} (data-flow slices of basic blocks) as the basic comparable unit. 
\texttt{Esh} uses SMT solver to verify function similarity, which makes it unscalable. 
As an improvement, the authors lift binaries to IR level and further create function-centric signatures~\cite{david2017similarity}.} 

\subsection{Machine Learning-based Code Similarity \zedit{Comparison}}

\noindent \textbf{Mono-architecture \zedit{solutions}.}
Recent research has demonstrated the usefulness of applying machine learning and deep learning 
techniques to code analysis~\cite{mokhov2014use,luo2016solminer,mou2016convolutional,
huo2016learning,white2017sorting,huo2017enhancing,nguyen2017exploring,
han2017learning,luo2016solminer}. 
White et al.\ further propose \texttt{DeepRepair} to detect the similarity between \emph{source code} 
fragments~\cite{white2017sorting}. Mou et al.\ {\color{black}introduce} a tree-based convolutional neural network based 
on program abstract syntax trees to detect similar \emph{source code} snippets~\cite{mou2016convolutional}. 
Huo et al.\ {\color{black}devise} NP-CNN~\cite{huo2016learning} and LS-CNN~\cite{huo2017enhancing} to identify buggy 
\emph{source code} files. 
{\color{black}\texttt{Asm2Vec}~\cite{dingasm2vec} produces a numeric vector for each function based on the PV-DM model~\cite{le2014distributed}. 
Similarity between two functions can be measured with the distances between the two corresponding representation vectors. 
$\alpha$\texttt{Diff}~\cite{liu2018alphadiff} characterizes a binary function using its code feature, invocation feature and module interactions feature, where the first category of  feature is learned from raw bytes with a DNN. However, this work only focuses on similarity detection between cross-version binaries.
} 
Zheng et al.~\cite{chua2017neural} independently {\color{black}propose} to use word embedding 
to represent instructions, but their word-embedding model does not address the issue of out-of-vocabulary 
(OOV) instructions, while handling OOV words has been a critical step in NLP \zedit{and is} resolved in our 
system (Section~\ref{sec:word-embed-train-dataset}); plus, their goal is to recover function signature 
from binaries of the same architecture, which is different from our cross-architecture code similarity 
comparison task.
Nguyen et al.\ {\color{black}develop} \texttt{API2VEC} for the API elements in \emph{source code} to measure code 
similarity~\cite{nguyen2017exploring}, which is not applicable if there are insufficient API calls. 

\noindent \textbf{Cross-architecture \zedit{solutions}.}
\texttt{Genius}~\cite{feng2016scalable} and \texttt{Gemini}~\cite{ccs2017graphembedding} are two \zedit{prior} state-of-the-art 
works on cross-architecture bug search. They make use of conventional machine learning and deep learning, 
respectively, to convert CFGs of \emph{functions} into vectors for similarity comparison.
BinGo~\cite{chandramohan2016bingo} {\color{black}introduces} a selective inlining technique to capture the function 
semantics and extracts \zedit{partial traces of various lengths} to model functions. 
{\color{black}However, all of these approaches} compare similarity between functions, and cannot handle code component 
similarity detection when only a \emph{part} of a function or code \emph{cross} 
function boundaries is reused in another program.

\noindent\textbf{\emph{Summary.}} Currently, no \zedit{solutions} are able to meet all these requirements: (a) working on binary code, (b) analyzing code of \emph{different architectures}, \zedit{(c) resolving the
code \emph{containment} problem.} 
This work fills the gap 
and proposes techniques for efficient cross-architecture binary code similarity comparison beyond function 
pairs. In addition, it is worth mentioning that many prior systems are built on basic block representation or comparison~\cite{gao2008binhunt,iBinhunt,luo2014semantics,pewny2015cross,feng2016scalable}; thus, 
\emph{they can benefit from our system which provides more precise basic block representation and efficient comparison.}

\section{\todo{Overview}}

Given a \emph{query} binary code component $\mathcal{Q}$, consisting of basic blocks whose 
relation can be represented in a control flow graph (CFG), we are interested in finding \zedit{programs,
from a large corpus of binary programs compiled for different 
architectures (e.g., x86 and ARM), that contain code components semantically equivalent or similar to $\mathcal{Q}$.} 
A code component here can be part of a function or contain multiple functions. 


\begin{figure*}[t]
\centering
\aaf
\graphicspath{figure/}
\includegraphics[width=0.92\textwidth]{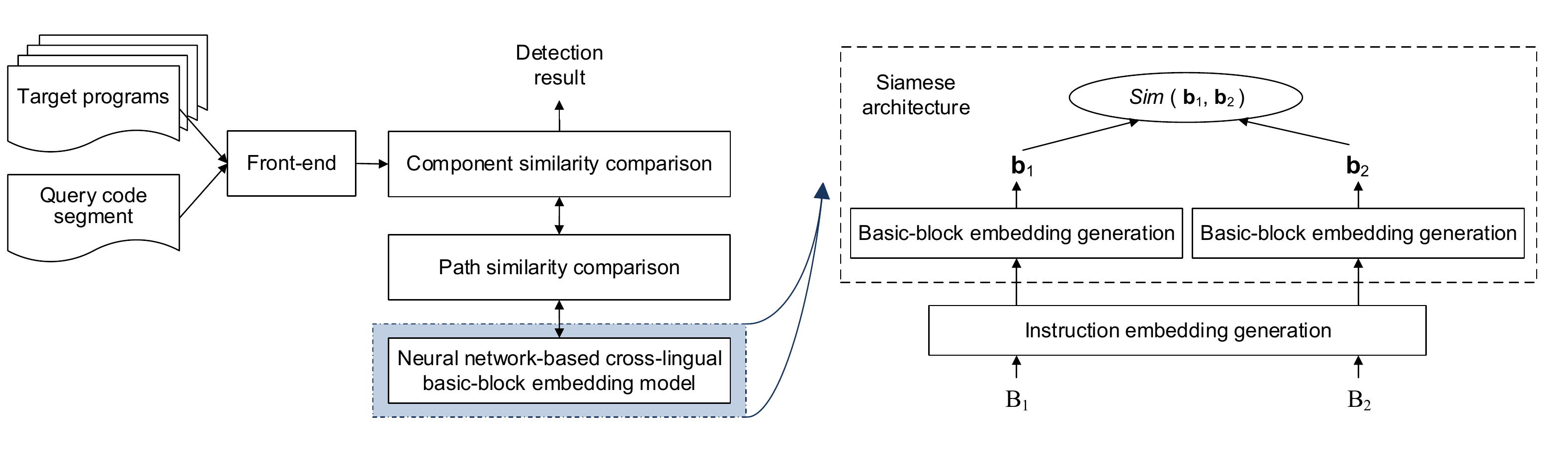}
\aaf\aaf\af
\caption{System architecture.}\label{fig:architecture}
\aaf\aaf
\end{figure*}

We \zedit{examine} code component semantics at three layers: basic blocks, CFG paths, 
and code components. The system architecture is shown in Figure~\ref{fig:architecture}. 
The inputs are the query code component and a set of target programs. The \emph{front-end} 
disassembles each binary and constructs its CFG. 
(1) To model the semantics of a basic block, we design the \emph{neural network-based cross-lingual 
basic-block embedding model} to represent a basic block as an embedding. The embeddings of all 
blocks are stored in a locality-sensitive hashing (LSH) database for efficient online search. 
(2) The \emph{path similarity comparison} component utilizes the {\color{black}LCS} \zedit{(Longest Common Subsequence)}
algorithm to compare the semantic similarity of two paths, one from the query code 
component and another from the target program, constructed based on the LCS dynamic programming 
algorithm with basic blocks as the sequence elements. The length of the common subsequence is 
then compared to the length of the path from the query code component. The ratio indicates the 
semantics of the query path as embedded in the target program. 
(3) The \emph{component similarity comparison} component explores multiple path pairs to collectively 
calculate a similarity score, indicating the likelihood of the query code component being reused 
in the target program. 

\noindent \textbf{Basic Block Similarity Detection.}
The key is to measure the similarity of two blocks, \emph{regardless of their target ISAs.} 
As shown in the right side of Figure~\ref{fig:architecture}, the neural network-based cross-lingual 
basic-block embedding model takes a pair of blocks as inputs,
and computes a similarity score $s\in[0, 1]$ as the output. The objective is that the more the two 
blocks are similar, the closer $s$ is to $1$, and the more the two blocks are dissimilar, the closer 
$s$ is to $0$. 
%
To achieve this, the model adopts a \emph{Siamese} architecture~\cite{bromley1994signature} with each side 
employing the {\color{black}LSTM}~\cite{hochreiter1997long}.
The Siamese architecture is a popular network architecture among tasks that involve finding similarity 
between two comparable things~\cite{bromley1994signature,chopra2005learning}. The {\color{black}LSTM} is capable 
of learning long range dependencies of a sequence. The two {\color{black}LSTM}s are trained \emph{jointly} to 
tolerate the cross-architecture syntactic variations.
\zedit{The model is trained using a large dataset, 
which contains a large number of basic block pairs with a similarity score as the label for each pair
(how to build the dataset is presented in 
Section~\ref{sec:block-embed-train-dataset})}.



A vector representation of an instruction and a basic block
is called an \emph{instruction embedding} and a \emph{block embedding}, respectively.
The block embedding model converts each block into \zedit{an embedding 
to facilitate comparison}. Specifically, three main steps are involved \zedit{in evaluating} 
the similarity of two blocks,
as shown \zedit{on} the right side of Figure~\ref{fig:architecture}. 
(1) Instruction embedding generation: given a block, each of its instructions is converted into an 
\emph{instruction embedding} using an instruction embedding matrix, which is learned via a neural 
network (Section~\ref{sec:instr-embedding-gen}). (2) Basic-block embedding generation:  \zedit{the embeddings of
instructions of each basic block are then} fed into a neural network 
to generate the \emph{block embedding} (Section~\ref{sec:bb-embedding-gen}). (3) Once the embeddings 
of two blocks have been obtained, their similarity can be \zedit{calculated} efficiently by 
measuring the distance between their block embeddings.

A prominent advantage of the model inherited from Neural Machine Translation is that it does 
not need to select features \emph{manually} when training the models; instead, as we will show later, 
the models \emph{automatically} {\color{black}learn} useful features during the training process. \zedit{Besides,  
prior state-of-the-art, \texttt{Genius}~\cite{feng2016scalable} and \texttt{Gemini}~\cite{ccs2017graphembedding},
which use \emph{manually} selected basic-block features, largely loses the information such as
the semantics of instructions and their dependencies. As a result, the precision of our approach outperforms 
theirs by large margin}. This is shown in our evaluation (Section~\ref{subsec:eval-comparison}).

\section{Instruction Embedding Generation} \label{sec:instr-embedding-gen}

An instruction includes an opcode (specifying the operation to be performed) and zero or 
more operands (specifying registers, memory locations, or literal data). For example, 
\texttt{mov eax, ebx} is an instruction where \texttt{mov} is an opcode and both \texttt{eax} 
and \texttt{ebx} are operands.\footnote{\zedit{Assembly code in this paper adopts the Intel syntax, i.e., 
\texttt{op dst, src(s)}.}} 
In NMT, words are usually converted into word embeddings to 
facilitate further processing. \zedit{Since we regard instructions as words, similarly we
represent instructions as \emph{instruction embeddings}.} 

Our notations use blackboard bold upper case to denote functions (e.g., $\mathbb{F}$), 
capital letters to denote basic blocks (e.g., \texttt{B}), bold upper case to represent 
matrices (e.g., \textbf{U}, \textbf{W}), bold lower case to represent vectors (e.g., 
\textbf{x}, $\textbf{y}_{i}$), and lower case to represent individual instructions in 
a basic block (e.g., $x_1$, $y_2$).

\subsection{Background: Word Embedding}
{\color{black}A unique aspect of NMT is its frequent use of word embeddings, 
which represent words in a high-dimensional space, to facilitate 
the further processing in neural networks.} In particular, a word embedding is to 
capture the contextual semantic meaning of the word; thus, \zedit{words with similar 
contexts have embeddings close to each other in the high-dimensional space}. Recently, 
a series of models~\cite{mikolov2013efficient,mikolov2013distributed,bengio2003neural} 
{\color{black}based on neural networks} have been proposed to learn high-quality word embeddings. 
Among these models, Mikolov's \emph{skip-gram} model is popular due to its efficiency and 
low memory usage~\cite{mikolov2013efficient}.

\begin{figure}
\aaf\af
\hspace{-21pt}
\graphicspath{figure/}
\includegraphics[scale=0.7]{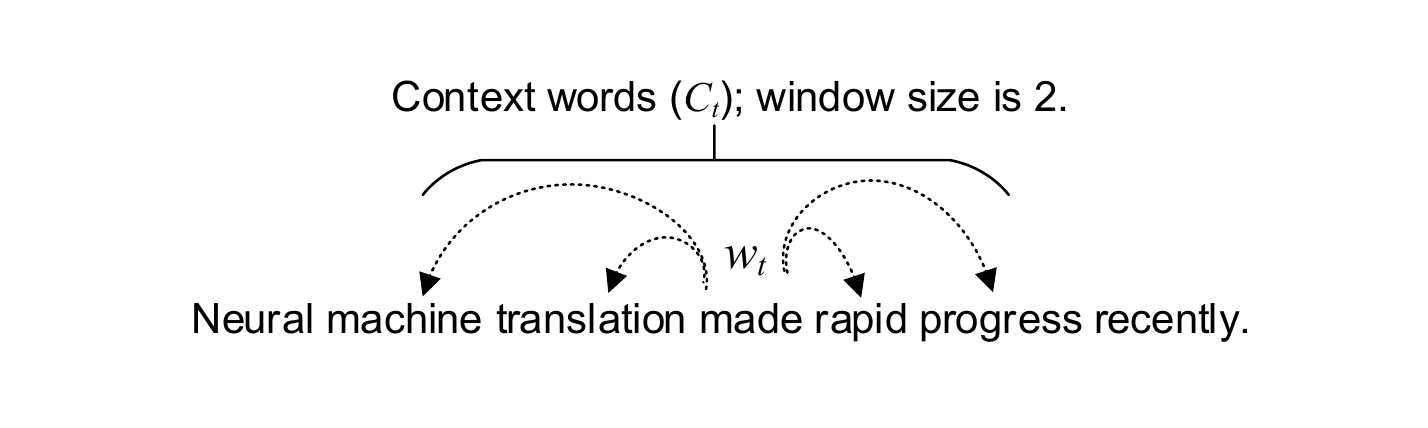}
\aaf\aaf\aaf\aaf\aaf
\caption{A sliding window used in skip-gram.}\label{fig:skip-gram}
\aaf\aaf\af
\end{figure}

The skip-gram model learns word embeddings by using a neural network. During training, 
a sliding window is employed on a text stream. In Figure~\ref{fig:skip-gram}, for
example, a window of size 2 is used, covering two words behind the current word 
and two words ahead. The model starts with a random vector for each word, and then 
gets trained when going over each sliding window. In each sliding window, the embedding 
of the current word, \textbf{w}$_{t}$, is used as the parameter vector of a \emph{softmax} 
function (Equation~\ref{equ_word}) that takes an arbitrary word $w_k$ as a training 
input and is trained to predict a probability of 1, if $w_k$ appears in the context 
$C_t$ (i.e., the sliding window) of $w_t$, and 0, otherwise. 

\aaf\af
\begin{equation} \label{equ_word}
P(w_k \in C_t|w_t) = \frac{\texttt{exp}(\textbf{w}_{t}^T \textbf{w}_{k})}{\sum_{w_i \in C_t} \texttt{exp}(\textbf{w}_{t}^T \textbf{w}_{i})}
\end{equation}
where \textbf{w}$_{k}$, \textbf{w}$_{t}$, and \textbf{w}$_{i}$ are the embeddings of 
words $w_k$, $w_t$, and $w_i$, respectively. 

Thus, given an arbitrary word $w_k$, its 
vector representation \textbf{w}$_{k}$ is used as a feature vector in the \emph{softmax} 
function parameterized by \textbf{w}$_{t}$. When trained on a sequence of $T$ words, the 
model uses stochastic gradient descent to maximize the log-likelihood objective $J(\texttt{w})$
as showed in Equation~\ref{equ_window}. 

\aaf\af
\begin{equation} \label{equ_window}
J(w)=\frac{1}{T}\sum_{t=1}^{T}\sum_{w_k \in C_t}(\texttt{log} \hspace{2pt} P(w_k|w_t))
\end{equation} 
\aaf\af

However, it would be very expensive to maximize $J(\texttt{w})$, because the denominator 
$\sum_{w_i \in C_t} \texttt{exp}(\textbf{w}_{t}^T \textbf{w}_{i})$ sums over all words 
$w_i$ in $C_t$. To minimize the computational cost, popular solutions are negative sampling 
and hierarchical softmax. We adopt the \emph{skip-gram with negative sampling model} 
(SGNS)~\cite{mikolov2013distributed}. After the model is trained on many sliding windows, 
the embeddings of each word become meaningful, yielding similar vectors for similar words. 
Due to its simple architecture and the use of the hierarchical softmax, the skip-gram model 
can be trained on a desktop machine at billions of words per hour. Plus, training the \zedit{model} 
is entirely \emph{unsupervised}.

\subsection{\todo{Challenges}}

Some unique challenges arise when learning instruction embeddings. 
First, in NMT, \zedit{a word embedding model
is usually trained once using large corpora, such as \emph{Wiki}, and then reused by other 
researchers}. However, we have to \emph{train an instruction embedding model 
from scratch}. 

Second, if a trained model is used to convert a word that has never appeared 
during training, the word is called an \emph{out-of-vocabulary} (OOV) word and the embedding 
generation for such words will fail. This is a well-known problem in NLP, and it
exacerbates significantly in our case, as constants, address offsets, labels, and strings 
are frequently used in instructions. How to deal with the OOV problem is a challenge.

\subsection{Building Training Dataset} \label{sec:word-embed-train-dataset}

Because we regard blocks as sentences, we use instructions of each block, 
called a \emph{Block-level Instruction Stream} (BIS) (Definition~\ref{def_path_stream}), 
to train the instruction embedding model. 

\newcommand*\concat{\mathbin{\|}}

\begin{definition}(Block-level Instruction Stream) \label{def_path_stream} 
Given a basic block \texttt{B}, consisting of a list of instructions.  
The \emph{block-level instruction stream} (BIS) of \texttt{B}, denoted as $\pi(\texttt{B})$,  
is defined as
\begin{equation*}
\pi(\texttt{B}) = (b_1,\cdots, b_n)
\end{equation*}
where $b_i$ is an instruction in \texttt{B}.
\end{definition}

\noindent \textbf{Preprocessing Training Data.}
%
%
To resolve the OOV problem, we propose to preprocess the instructions in the
training dataset using the following rules: 
(1) The number constant values are replaced with 0, and the minus signs are preserved.
(2) The string literals are replaced with \verb|<STR>|.
(3) The function names are replaced with \verb|FOO|.
(4) Other symbol constants are replaced with \verb|<TAG>|.
Take the following code snippets as an example: the left code snippet shows the 
original assembly code, and the right one is the preprocessed result. 

\begin{center}
{\small
\aaf
\begin{tabular}{l@{\qquad}|@{\qquad}l}
\texttt{\textbf{MOVL} \%ESI, \$.L.STR.31} & \texttt{\textbf{MOVL} ESI, <STR>}\\
\texttt{\textbf{MOVL} \%EDX, \$3} & \texttt{\textbf{MOVL} EDX, 0} \\
\texttt{\textbf{MOVQ} \%RDI, \%RAX} & \texttt{\textbf{MOVQ} RDI, RAX}\\
\texttt{\textbf{CALLQ} STRNCMP} & \texttt{\textbf{CALLQ} FOO} \\
\texttt{\textbf{TESTL} \%EAX, \%EAX} & \texttt{\textbf{TESTL} EAX, EAX}\\
\texttt{\textbf{JE} .LBB0\_5} & \texttt{\textbf{JE} <TAG>}
\end{tabular}
\af}
\end{center}

Note that the same preprocessing rules are applied to instructions before generating their
embeddings. This way, we can significantly reduce the OOV cases. Our evaluation result
(Section~\ref{sec:vocabulary-eval}) 
shows that, after a large number of preprocessed 
instructions 
are collected to train the model, we encounter very few OOV cases in the
later testing phase. This means the trained model is readily \emph{reusable} 
for other researchers. Moreover, semantically similar instructions indeed have embeddings
that are close to each other, as predicted.

\subsection{Training Instruction Embedding Model} \label{subsec:learningstage}

\begin{figure}
\centering
\vspace*{3pt}
\graphicspath{figure/}
\includegraphics[scale=0.67]{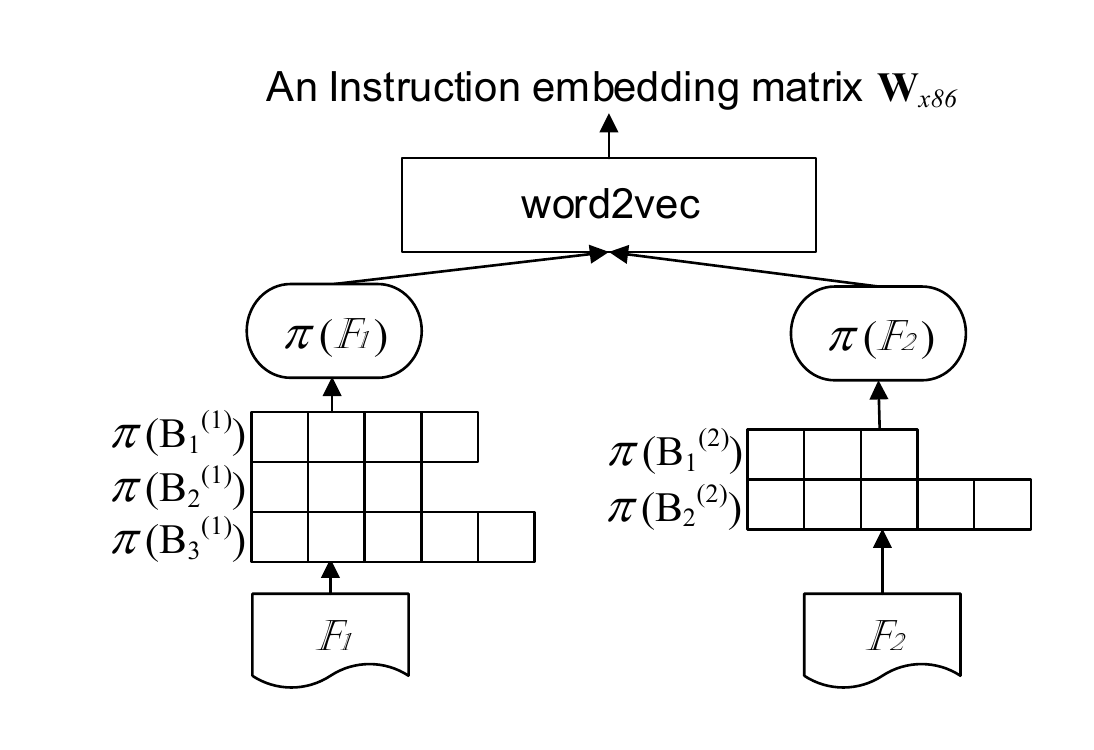}
\af
\caption{Learning instruction embeddings for x86. $\pi(\texttt{B}_i^{(j)})$ represents the 
$i$-th block-level instruction stream (BIS) in the function $\mathbb{F}_j$. Each square 
in a BIS represents an instruction.}
\label{fig:wordvec}
\aaf\aaf
\end{figure}

\zedit{We adopt the skip-gram negative sampling model as implemented in \texttt{word2vec}~\cite{mikolov2013efficient} 
to build our instruction embedding model.}
As an example,
Figure~\ref{fig:wordvec} shows the process of training the model for the x86 architecture. 
\zedit{For each architecture, an architecture-specific model is trained using the functions in our dataset containing
binaries of that architecture.} 
Each function is parsed to generate the corresponding 
Block-level Instruction Streams (BISs), which are fed, BIS by BIS, into the model for training. 
The training result is {\color{black}an embedding matrix containing the numerical representation} of each instruction.

{\color{black}The resultant instruction embedding matrix is denoted by $\textbf{W} \in \mathbb{R}^{d^{e} \times V}$,} 
where $d^{e}$ is the dimensionality of the instruction embedding selected by users (how 
to select $d^{e}$ is discussed in Section~\ref{subsec:eval-hyper}) and $V$ is the number 
of distinct instructions in the vocabulary. \zedit{The \emph{i-th} column of $\textbf{W}$
corresponds to the instruction embedding of the \emph{i-th} instruction in the vocabulary} (all 
distinct instructions form a vocabulary).



\section{Block Embedding Generation}  \label{sec:bb-embedding-gen}


A straightforward attempt for generating the embedding of a basic block is to simply 
compose (e.g., summing up) all embeddings of the instruction in the basic block. 
However, this processing 
cannot handle the cross-architecture differences, as instructions that stem from the 
same source code but \zedit{belong to} different architectures may have very different embeddings.
This is verified in our evaluation (Section~\ref{subsec:eval-instru-embed}).

\begin{figure}
\hspace{15pt}
\includegraphics[scale=0.58]{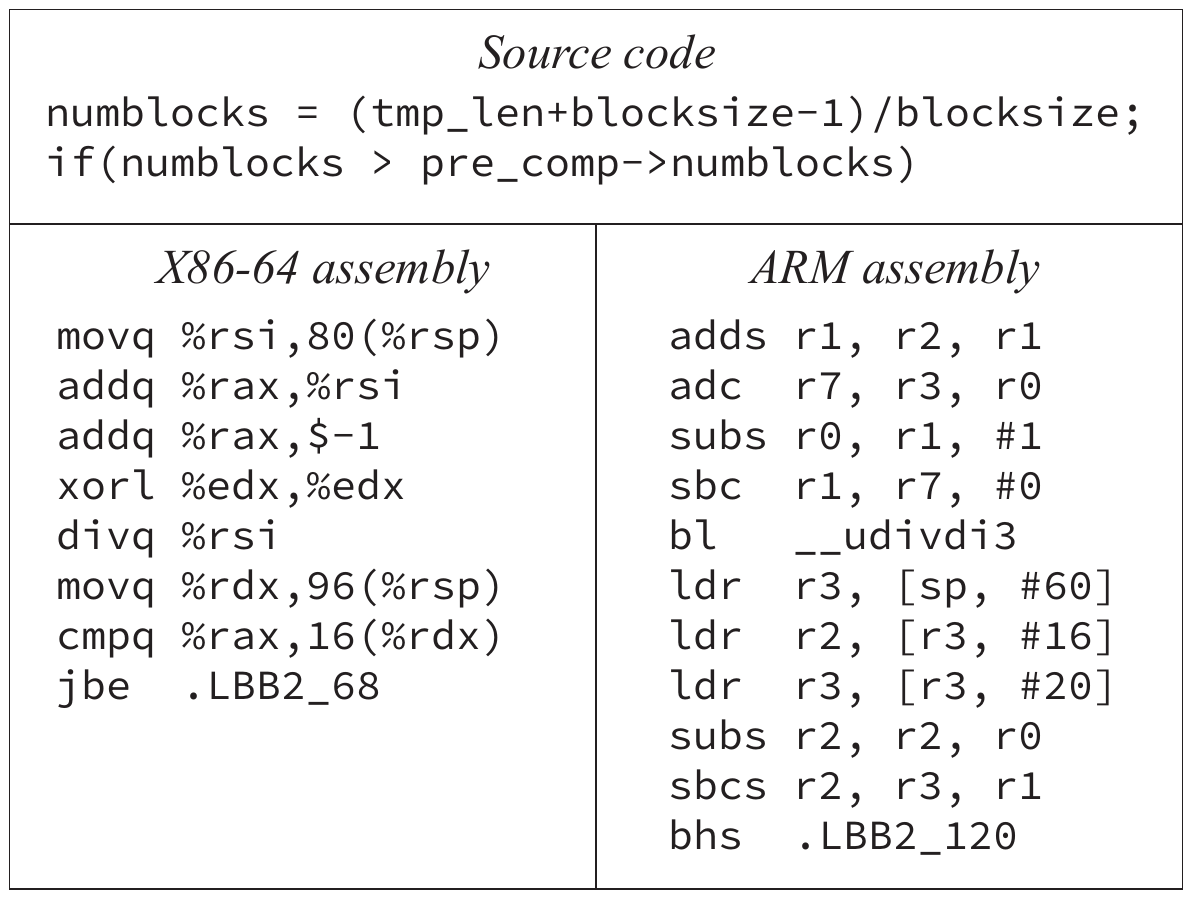}
\af
\caption{C source of a basic block from \texttt{ec\_mult.c} in \texttt{OpenSSL} and the 
assembly code for two architectures.}\label{fig:assembly}
\aaf\aaf\af
\end{figure}

\zedit{Figure~\ref{fig:assembly} shows a code snippet (containing 
one basic block) that has been compiled targeting two different architectures,  x86-64 and ARM. 
While the two pieces of binary code are semantically equivalent, they look very different due 
to different instructions sets, CPU registers, and 
memory addressing modes. The basic block embedding generation should be able to handle
such syntactic variation.}

\subsection{Background: {\color{black}LSTM} in NLP} \label{subsec:rnn-lstm}

RNN is a type of deep neural network that has been 
successfully applied to \zedit{converting} word embeddings of a sentence to a sentence 
embedding~\cite{chung2014empirical,kalchbrenner2014convolutional}.
{\color{black}As a special kind of RNN, LSTM is developed to address the difficulty of capturing long term 
memory in the basic RNN.} 
A limit of 500 words for the sentence length is often used in practice, 
and a basic block usually contains less than 500 instructions.

In text analysis, {\color{black}LSTM} treats a sentence as a sequence of words with internal 
structures, i.e., word dependencies. It encodes the semantic vector of 
a sentence incrementally, left-to-right and word-by-word. 
At each time step, a new word is encoded and the word dependencies 
embedded in the vector are ``updated''. 
When the process reaches the end of the sentence, the semantic vector has embedded all 
the words and their dependencies, and hence, can be viewed as a feature representation 
of the whole sentence. That semantic vector is the sentence embedding.

\subsection{Cross-lingual Basic-block Embedding Model Architecture} \label{{sec:solution}}

\begin{figure}
\centering
\graphicspath{figure/}
\includegraphics[scale=0.35]{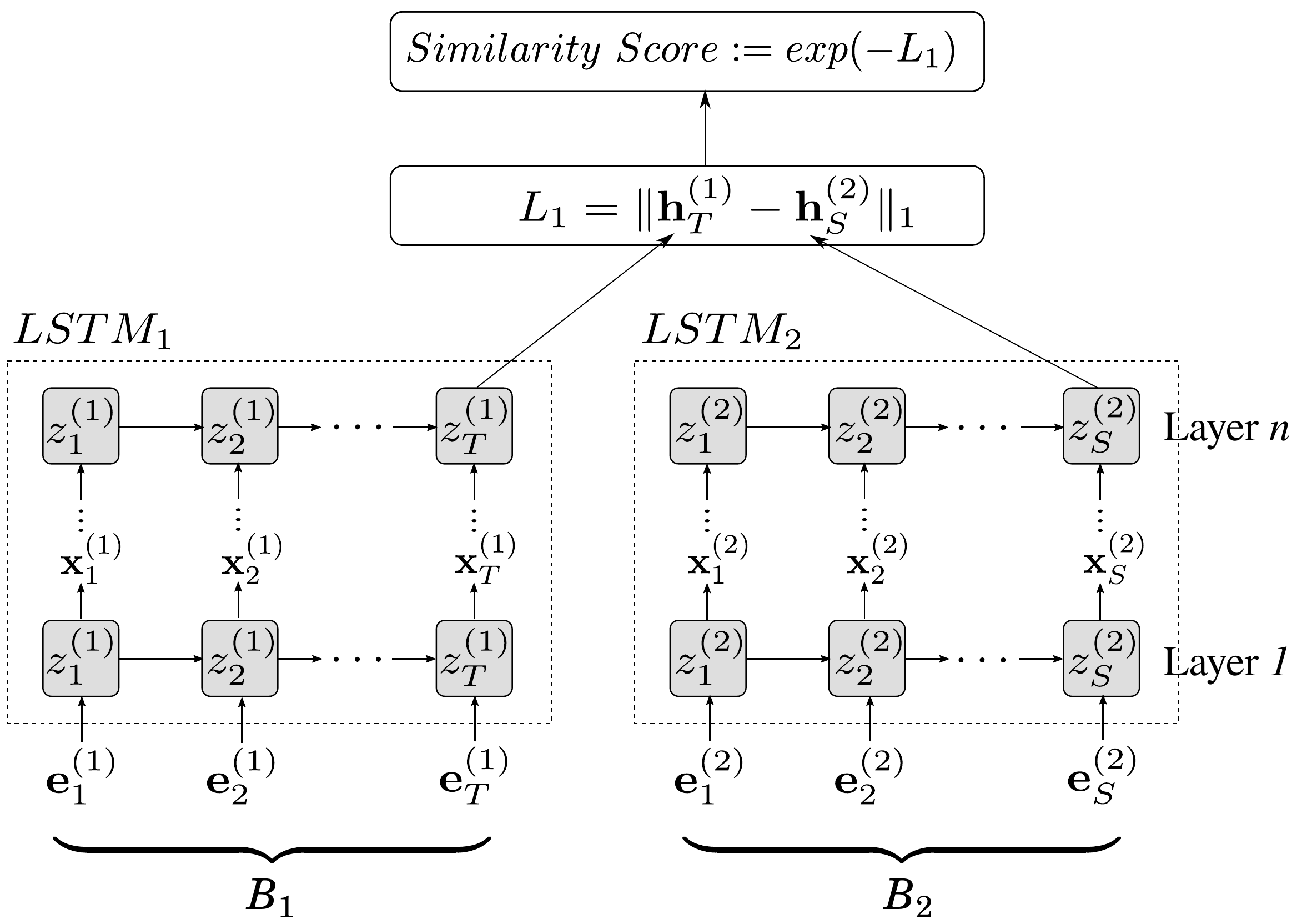}
\af
\caption{Neural network-based basic-block embedding model. {\color{black}Each shaded box is an LSTM cell.}}\label{fig:funvec}
\aaf\aaf
\end{figure}

Inspired by the NMT model that compares the similarity of sentences of different languages, 
we design a \emph{neural network-based cross-lingual basic-block embedding model} to compare 
the semantics similarity of basic blocks of different ISAs. As showed in Figure~\ref{fig:funvec},
we design our model as a \emph{Siamese architecture}~\cite{bromley1994signature} with each side 
employing the \emph{identical} {\color{black}LSTM}. 
Our objective is to \emph{make the embeddings for blocks of similar semantics as 
close as possible, and meanwhile, to make blocks of different semantics as far apart as 
possible}. A Siamese architecture takes the embeddings of instructions in two blocks, 
$\texttt{B}_1$ and $\texttt{B}_2$, as inputs, and produces the similarity score as an output. 
This model is trained with only supervision on a basic-block pair as input and the ground truth 
$\chi(\texttt{B}_1, \texttt{B}_2)$ as an output \emph{without relying on any manually selected features}. 

For embedding generation, each LSTM cell sequentially takes an input (for the first layer 
the input is an instruction embedding) at each time step, accumulating and passing 
increasingly richer information. When the last instruction embedding is reached, the 
last LSTM cell at the last layer provides a  semantic representation of the basic block, 
i.e., a block embedding. Finally, the similarity of the two basic blocks is measured as 
the distance of the two block embeddings.

\noindent \textbf{Detailed Process.}
The inputs are two blocks, $\texttt{B}_1$ and $\texttt{B}_2$, represented as a sequence 
of instruction embeddings, ($\textbf{e}_1^{(1)}, \cdots, \textbf{e}_T^{(1)}$), and 
($\textbf{e}_1^{(2)}, \cdots, \textbf{e}_S^{(2)}$), respectively. Note that the sequences 
may be of different lengths, i.e., $|T| \neq |S|$, and the sequence lengths can vary from 
example to example; both are handled by the model. {\color{black}An LSTM cell analyzes an input vector coming from either the input 
embeddings or the precedent step and updates its hidden state at each 
time step}. Each cell contains four components (which 
are real-valued vectors): a \emph{memory state} $c$, an \emph{output gate} $o$ determining 
how the memory state affects other units, and an \emph{input gate} $i$ (and a \emph{forget gate} 
$f$, resp.) that controls what  gets stored in (and omitted from, resp.) memory. 
For example, an {\color{black}LSTM cell} at the first layer in {\color{black}LSTM$_1$} updates 
its hidden state at the time step $t$ via Equations~\ref{equ:lstm-beg}--\ref{equ:lstm-end}:

\aaf\af
\begin{equation} \label{equ:lstm-beg}
\footnotesize
\textbf{i}_t^{(1)} = \texttt{sigmoid}(\textbf{W}_i \textbf{e}_t^{(1)} + \textbf{U}_i \textbf{x}_{t-1}^{(1)} + \textbf{v}_i)
\end{equation}
\begin{equation} 
\footnotesize
\textbf{f}_t^{(1)} = \texttt{sigmoid}(\textbf{W}_f \textbf{e}_t^{(1)} + \textbf{U}_f \textbf{x}_{t-1}^{(1)} + \textbf{v}_f)
\end{equation}
\begin{equation}
\footnotesize
\widetilde{\textbf{c}_t}^{(1)} = \texttt{tanh}(\textbf{W}_c \textbf{e}_t^{(1)} + \textbf{U}_c \textbf{x}_{t-1}^{(1)} + \textbf{v}_c)
\end{equation}
\begin{equation}
\footnotesize
\textbf{c}_t^{(1)} = \textbf{i}_t^{(1)} {\color{black}\odot} \widetilde{\textbf{c}_t}^{(1)} + \textbf{f}_t^{(1)} {\color{black}\odot} \widetilde{\textbf{c}_t}^{(1)} 
\end{equation}
\begin{equation}
\footnotesize
\textbf{o}_t^{(1)} = \texttt{sigmoid}(\textbf{W}_o \textbf{e}_t^{(1)} + \textbf{U}_o \textbf{x}_{t-1}^{(1)} + \textbf{v}_o)
\end{equation}
\begin{equation} \label{equ:lstm-end}
\footnotesize
\textbf{x}_t^{(1)} = \textbf{o}_t^{(1)} {\color{black}\odot} \texttt{tanh}(\textbf{c}_t^{(1)})
\end{equation}
where {\color{black}$\odot$} denotes Hadamard (element-wise) product; $\textbf{W}_i$, $\textbf{W}_f$, 
$\textbf{W}_c$, $\textbf{W}_o$, $\textbf{U}_i$, $\textbf{U}_f$, $\textbf{U}_c$, $\textbf{U}_o$ 
are weight matrices; and $\textbf{v}_i$, $\textbf{v}_f$, $\textbf{v}_c$, $\textbf{v}_o$ 
are bias vectors; they are learned during training. 
The reader is referred to~\cite{hochreiter1997long} for more details.

At the last time step $T$, the last hidden state at the last layer provides a vector 
$\textbf{h}_T^{(1)}$ (resp. $\textbf{h}_S^{(2)}$ ), which is the embedding of $\texttt{B}_1$ 
(resp. $\texttt{B}_2$). We use the Manhattan distance ($\in [0, 1]$) {\color{black}which is suggested by~\cite{mueller2016siamese} to} measure the similarity 
of $\texttt{B}_1$ and $\texttt{B}_2$ as showed in Equation~\ref{equ:similar}:

\aaf\af
\begin{equation} \label{equ:similar}
\texttt{Sim}(\texttt{B}_1, \texttt{B}_2) = \texttt{exp}(-\concat \textbf{h}_{T}^{(1)} - \textbf{h}_{S}^{(2)}\concat _1)
\aaf
\end{equation}

To train the network parameters, we use stochastic gradient descent (SGD) to minimize 
the \emph{loss} function: 

\aaf\aaf
\begin{equation} \label{equ:loss}
\min_{\textbf{W}_i, \textbf{W}_f, ..., \textbf{v}_o} \sum_{i=1}^{N} (y_i - \texttt{Sim}(\texttt{B}_1^i, \texttt{B}_2^i))^2
\end{equation}
where $y_i$ is the similarity ground truth of the pair $<\texttt{B}_1^i, \texttt{B}_2^i>$, 
and $N$ the number of basic block pairs in the training dataset. 

In the end, once the Area Under the Curve (AUC) value converges, the training process terminates, 
and the trained cross-lingual basic-block embedding model is capable of encoding an input 
binary block to an embedding capturing the semantics information of the block that is suitable 
for similarity detection.

\subsection{\todo{Challenges}}

There are two main challenges for learning block embeddings. First, in order to train, validate 
and test the basic-block embedding model, a large dataset containing labeled similar and 
dissimilar block pairs is needed. Unlike prior work~\cite{ccs2017graphembedding} that builds 
the dataset of similar and dissimilar \emph{function} pairs by using the function names 
to establish the ground truth about the \emph{function} similarity, it is very challenging to  
establish the ground truth for basic blocks because: (a) no name is available to indicate whether two basic 
blocks are similar or not, and (b) even if two basic blocks have been compiled from two 
pieces of code, they may happen to be equivalent or similar, and therefore, it would be incorrect to label 
them as dissimilar.

Second, many hyperparameters need to be determined to maximize the model performance. 
The parameter values selected for NMT are not necessarily applicable to our model, and need to be 
comprehensively examined (Section~\ref{subsec:eval-hyper}).

\subsection{Building Dataset} \label{sec:block-embed-train-dataset}


\subsubsection{Generating Similar Basic-Block Pairs}

\begin{figure}
\centering
\vspace{5pt}
\includegraphics[scale=0.5]{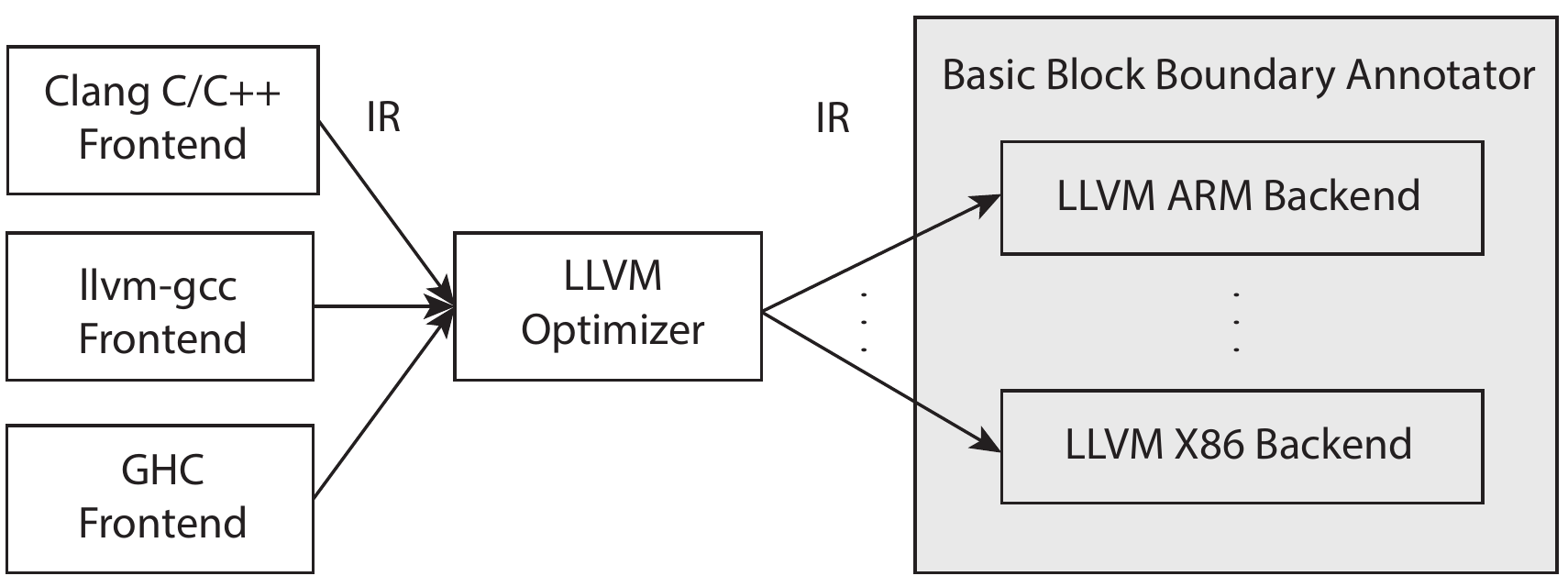}
\caption{LLVM architecture. The basic-block boundary
annotator is added into the backends of different architectures.}\label{fig:llvm}
\aaf\aaf
\end{figure}

We consider two basic blocks of different ISAs that have been compiled from the same piece of 
source code as equivalent. To establish the ground truth about the block similarity, 
we modify the \emph{backends} of various architectures in the LLVM compiler. 
As shown in Figure~\ref{fig:llvm}, 
the LLVM compiler consists of various frontends (that compile source code 
into a uniform Intermediate Representation (IR)), the middle-end optimizer, and various architecture-dependent 
backends (that generate the corresponding assembly code). We modify the backends to add the basic-block 
boundary annotator, which not only clearly marks the boundaries of blocks, but also annotates 
a unique ID for each generated assembly block in a way that \emph{all assembly blocks 
compiled from the same IR block (i.e., the same piece of source code), regardless of their architectures, 
will obtain the same ID}.

To this end, we collect various open-sourced software projects, and feed 
the source code into the modified LLVM compiler to generate a large number of basic blocks
for different architectures. After preprocessing (Section~\ref{sec:word-embed-train-dataset})
and data deduplication, for each basic \zedit{block \texttt{B$^{x86}$}, the basic block} 
\texttt{B$^{ARM}$} with the same ID is sampled to construct one training example 
<\texttt{B$^{x86}$}, \texttt{B$^{ARM}$}, 1>. By continually 
sampling, we can collect a large number of similar basic-block pairs.

\subsubsection{Generating Dissimilar Basic-Block Pairs}

While two basic blocks with the same ID are always semantically equivalent, two blocks 
with different IDs may not necessarily be dissimilar,
as they may happen to be be equivalent. 


To address this issue, we \zedit{make use of} \emph{$n$-gram} to measure the text similarity between \zedit{two basic blocks compiled for the \emph{same} architecture at the same optimization level.} 
A low text similarity score indicates 
two basic blocks are dissimilar.  Next, assume a block \texttt{B$_{1}^{ARM}$} of
ARM is \emph{equivalent} to a block \texttt{B$_{1}^{x86}$} of x86 (they have the same ID);  
and another block \texttt{B$_{2}^{x86}$} of x86 is \emph{dissimilar} to \texttt{B$_{1}^{x86}$} 
according to the $n$-gram similarity comparison. Then, the two blocks, \texttt{B$_{1}^{ARM}$} and \texttt{B$_{2}^{x86}$}, 
are regarded as dissimilar, and the 
instance <\texttt{B$_{1}^{ARM}$}, \texttt{B$_{2}^{x86}$}, 0> 
is added to the dataset. Our experiments set $n$ as 4 and the similarity threshold as 0.5; that is, 
if two blocks, through this procedure, have a similarity score smaller than 0.5, they are labeled as 
dissimilar. This way, we can obtain a large number of dissimilar basic-block pairs across  
architectures.

\section{Path/Code Component Similarity Comparison}

\todo{Detecting similar code components is an important problem. Existing work
either can only  work on a \emph{single} architecture~\cite{wang2009behavior,
gao2008binhunt,luo2014semantics,iBinhunt,tamada2004dynamic,schuler2007dynamic,jhi2011value}, or can compare \emph{a pair of functions} across architectures~\cite{pewny2015cross,
eschweiler2016discovre,feng2016scalable,ccs2017graphembedding}. 
However, as a critical code part may be {\color{black}inserted} inside a function
to avoid detection~\cite{jiang2007deckard,jhi2011value,luo2014semantics}, how to resolve
the cross-archite code containment problem is a \emph{new} and more challenging problem. }

\todo{We propose to decompose the CFG of the query code component $\mathcal{Q}$ into multiple paths. 
For each path from $\mathcal{Q}$, we compare it to many paths from the target program $\mathcal{T}$, 
to calculate a path similarity score by adopting the \emph{Longest Common Subsequence} (LCS) 
dynamic programming algorithm with basic blocks as sequence elements. By trying more than one 
path, we can use the path similarity scores collectively to detect whether a component in 
$\mathcal{T}$ is similar to $\mathcal{Q}$.}

\subsection{Path Similarity Comparison} \label{sec:path-comp}

A linearly independent path is a path that introduces at least one new node (i.e., basic block) 
that is not included in any previous linearly independent paths~\cite{watson1996}. 
Once the starting block of $\mathcal{Q}$ and several candidate starting blocks of $\mathcal{T}$ 
are identified (presented in Section~\ref{sec:com-comp}), the next step is to explore paths to calculate a path 
similarity score. For $\mathcal{Q}$, we select a set of linearly independent paths from the 
starting block. We first unroll each loop in $\mathcal{Q}$ once, and adopt the Depth 
First Search algorithm to find a set of linearly independent paths.


For each linearly independent path of $\mathcal{Q}$, we need to find the highest 
similarity score between the query path and the many paths of $\mathcal{T}$. To this end, 
we apply a recently proposed code similarity comparison approach, called \texttt{CoP}~\cite{luo2014semantics} 
(it is powerful for handling many types of obfuscations but can only handle code components of 
the same architecture). \texttt{CoP} combines the {\color{black}LCS} algorithm and 
basic-block similarity comparison to compute the LCS of semantically equivalent basic blocks (SEBB). 
However, \texttt{CoP}'s basic-block similarity comparison relies on symbolic execution and theorem 
proving, which is very computationally expensive~\cite{luo2017system}. On the contrary, our work adopts techniques in NMT 
to \emph{significantly speed up basic-block similarity comparison}, and hence is much more scalable 
for analyzing large codebases.

Here we briefly introduce how \texttt{CoP} applies LCS to detect path similarity. 
It adopts breadth-first search in the \emph{inter-procedural} CFG of the target program 
$\mathcal{T}$, 
combined with the LCS dynamic programming to compute the highest score of the LCS of SEBB. 
For each step in the breadth-first dynamic programming algorithm, the LCS is kept as the 
``longest path'' computed so far for a block in the query path. The LCS score of the 
last block in the query path is the highest LCS score, and is used to compute a path 
similarity score. 
Definition~\ref{def_pathscore_task3} gives a high-level description of a path similarity score. 

\begin{definition}{(Path Similarity Score)}
Given a linearly independent path $\mathcal{P}$ from the query code component, 
and a target program  $\mathcal{T}$. Let $\Gamma =\{\mathcal{P}_1^{t}, \ldots, \mathcal{P}_n^{t}\}$
be all of the linearly independent paths of $\mathcal{T}$, and 
$|\mathrm{LCS}(\mathcal{P}, \mathcal{P}_i^{t})|$ be the length of the LCS of SEBB 
between $\mathcal{P}$ and $\mathcal{P}_i^{t}$, $\mathcal{P}_i^{t} \in \Gamma$. Then, the path 
similarity score for $\mathcal{P}$ is defined as

\aaf\af
\begin{equation*}
\psi(\mathcal{P},T)=\frac{\max \nolimits_{\mathcal{P}_i^{t} \in \Gamma} |\mathrm{LCS}(\mathcal{P}, \mathcal{P}_i^{t})|}{|\mathcal{P}|}
\end{equation*}
\label{def_pathscore_task3}
\end{definition}
\aaf

\subsection{Component Similarity Comparison} \label{sec:com-comp}

\noindent \textbf{Challenge.}
\emph{The location that the code component gets embedded into the containing target program is unknown}, 
and it is possible for it to be inserted into the middle of a function. It is important to determine the correct 
starting points so that the path exploration is not misled to irrelevant code parts of the target 
program. This is \emph{a unique challenge} compared to function-level code similarity comparison.

\noindent \textbf{Idea.}
We look for the starting blocks in the manner as follows. First, the embeddings of all basic blocks of the target 
program $\mathcal{T}$ are stored in an locality-sensitive hashing database for efficient online search. Next, 
we begin with the first basic block in the query code component $\mathcal{Q}$ as the starting block, and search
in the database to find a semantically equivalent basic block (SEBB) from the target program $\mathcal{T}$. If 
we find one or several SEBBs, we proceed with the path exploration (Section~\ref{sec:path-comp}) 
on each of them. Otherwise, we choose another block from  $\mathcal{Q}$ as the starting 
block~\cite{luo2014semantics}, and repeat the process until the last block of  $\mathcal{Q}$ is checked.

\noindent \textbf{Component similarity score.}
We select a set of linearly independent paths from  $\mathcal{Q}$, and compute a path 
similarity score for each linearly independent path. Next, we assign a weight to each path similarity 
score according to the length of the corresponding query path. The final component similarity score 
is the weighted average score. 

\noindent \emph{\textbf{Summary.}}
By integrating our cross-lingual basic-block embedding model with an existing 
approach~\cite{luo2014semantics}, we have come up with an effective and efficient solution 
to cross-architecture code-component similarity comparison. Moreover, it demonstrates how 
the efficient, precise and scalable basic-block embedding model can benefit  \zedit{many other}
systems~\cite{gao2008binhunt,luo2014semantics,iBinhunt} that rely on basic-block similarity comparison.

\section{Evaluation} \label{sec:evaluation}

We evaluate \tool\ \zedit{in terms of} its accuracy, efficiency, and scalability. 
First, we describe the \zedit{experimental settings} (Section~\ref{subsec:eval-setup}) and 
discuss the datasets used in our evaluation (Section~\ref{subsec:eval-dataset}). 
Next, we examine the impact of preprocessing on out-of-vocabulary instructions (Section~\ref{sec:vocabulary-eval})
and the {\color{black}quality}  of the instruction embedding model (Section~\ref{subsec:eval-instru-embed}).
We then evaluate whether \bbtool\ can successfully detect the similarity of blocks 
compiled for different architectures (Problem \textbf{I}). We evaluate its accuracy and efficiency 
(Sections~\ref{subsec:eval-accuracy} and \ref{subsec:eval-efficiency}), and discuss 
hyperparameter selection (Section~\ref{subsec:eval-hyper}). We also compare it with 
a machine learning-based basic-block comparison approach that uses a set of manually 
selected features~\cite{feng2016scalable,ccs2017graphembedding} (Section~\ref{subsec:eval-comparison}). 
Finally, we present three real-world case studies demonstrating how \cctool\ can be applied 
for cross-architecture code component search and cryptographic function search under realistic 
conditions (Problem \textbf{II}) in Section~\ref{subsec:bugsearch}.

\subsection{\zedit{Experimental Settings}} \label{subsec:eval-setup}

We adopt \texttt{word2vec}~\cite{mikolov2013efficient} to learn instruction embeddings, 
and implemented our cross-lingual basic-block embedding model in Python using the 
\texttt{Keras}~\cite{chollet2015keras} platform with \texttt{TensorFlow}~\cite{abadi2016tensorflow} 
as backend. \texttt{Keras} provides a large number of high-level neural network APIs 
and can run on top of \texttt{TensorFlow}. Like the work \texttt{CoP}~\cite{luo2014semantics}, 
we require that the selected linearly independent paths cover at least 80\% of the basic blocks 
in each query code component; the largest number of the selected linearly independent paths 
in our evaluation is 47. \cctool\ (the LCS algorithm with path exploration) is implemented 
in the BAP framework~\cite{bap} which constructs CFGs and call graph and builds the 
inter-procedural CFG. \cctool\ queries the block embeddings (computed by \bbtool) stored 
in {\color{black}an} LSH database. The experiments \zedit{are} performed on a computer running the Ubuntu 14.04 
operating system with a 64-bit 2.7 GHz Intel\textsuperscript{\textregistered} Core\textsuperscript{(TM)} 
i7 CPU and 32 GB RAM \zedit{\emph{without GPUs}. The training and testing are expected to
be significantly accelerated if GPUs are used.}

\subsection{Dataset} \label{subsec:eval-dataset}

\begin{table*}[!h]
\small
\hspace{-6pt}
\caption{The number of basic-block pairs in the training, validation and testing datasets.}\label{tab:dataset}
\aaf
\renewcommand{\arraystretch}{1.1}
\scalebox{0.97}{
\setlength{\belowcaptionskip}{10pt}
\centering
\begin{tabular}{|c|c|c|c|c|c|c|c|c|c||c|c|c|}
\hline
\multirow{2}*{ } & \multicolumn{3}{c|}{Training} & \multicolumn{3}{c|}{Validation}  & \multicolumn{3}{c||}{Testing} & \multicolumn{3}{c|}{\textbf{Total}}   \\
\cline{2-13}
                 & Sim. & Dissim. & Total  & Sim.   &  Dissim. & Total  & Sim. &  Dissim. & Total  &  \textbf{Sim.} &  \textbf{Dissim.} & \textbf{Total}  \\ \hline
O1               & 35,416   & 35,223   & 70,639 & 3,902      & 3,946    & 7,848  & 4,368    & 4,354    & 8,722  & 43,686 & 43,523& 87,209  \\ 
O2               & 45,461   & 45,278   & 90,739 & 5,013      & 5,069    & 10,082 & 5,608    & 5,590    & 11,198 & 56,082 & 55,937 & 112,019 \\ 
O3               & 48,613   & 48,472   & 97,085 & 5,390      & 5,397    & 10,787 & 6,000    & 5,988    & 11,988 & 60,003 & 59,857 & 119,860 \\ 
Cross-opts       & 34,118   & 33,920   & 68,038 & 3,809      & 3,750    & 7,559  & 4,554    & 4,404    & 8,958  & 42,481 & 42,074 & 84,555  \\ 
\hhline{|=|=|=|=|=|=|=|=|=|=||=|=|=|}
\textbf{Total}   & 163,608  & 162,893   &   326,501     & 18,114 &     18,162   &  36,276      &  20,530   &  20336   &  40,866   & 202,252 & 201,391 & 403,643    \\
\hline
\end{tabular}
}
\af
\end{table*}

We first describe the dataset (\textbf{Dataset I}), as shown in Table~\ref{tab:dataset}, 
used to evaluate the cross-lingual basic-block embedding 
model (\bbtool). All basic-block pairs in the dataset are labeled with the similarity ground truth. 
In particular, we prepare this dataset using \texttt{OpenSSL} (v1.1.1-pre1) and four popular Linux 
packages, including \texttt{coreutils} (v8.29), \texttt{findutils} (v4.6.0), \texttt{diffutils} (v3.6),
and \texttt{binutils} (v2.30). We use two architectures (x86-64 and ARM) and \texttt{clang} (v6.0.0) 
with {\color{black}three} different optimization levels (O1-O3) to compile each program. 
In total, we obtain 437,104 basic blocks for x86, and 393,529 basic blocks for ARM.

\zedit{We follow the approach described in  Section~\ref{sec:block-embed-train-dataset} to
generate similar/dissimilar basic-block pairs.} Totally, we generate 202,252 
similar basic-block pairs (\emph{one compiled from x86 and another from ARM}; as shown in 
the 11th column of Table~\ref{tab:dataset}), where 43,686 pairs, 56,082 pairs, 60,003 pairs, 
and 42,481 pairs are compiled using O1, O2, O3, and different optimization levels, respectively.  
Similarly, we generate 201,391 dissimilar basic-block pairs (as shown in the 12th column of 
Table~\ref{tab:dataset}), where 43,523 pairs, 55,937 pairs, 59,857 pairs, and 42,074 pairs 
are compiled using O1, O2, O3, and different optimization levels, respectively.

\subsection{Evaluation on Out-Of-Vocabulary Instructions} \label{sec:vocabulary-eval}

As pre-processing is applied to \zedit{addressing} the issue of out-of-vocabulary (OOV) instructions
(Section~\ref{sec:word-embed-train-dataset}), we evaluate its impact, and seek to understand:
a) how the vocabulary size (the number of columns in the instruction embedding matrix)
grows with or without pre-processing, and b) the number of OOV cases in later instruction 
embedding generation. 

\begin{figure}
\begin{minipage}{0.5\textwidth}
\centering
\includegraphics[scale=0.38]{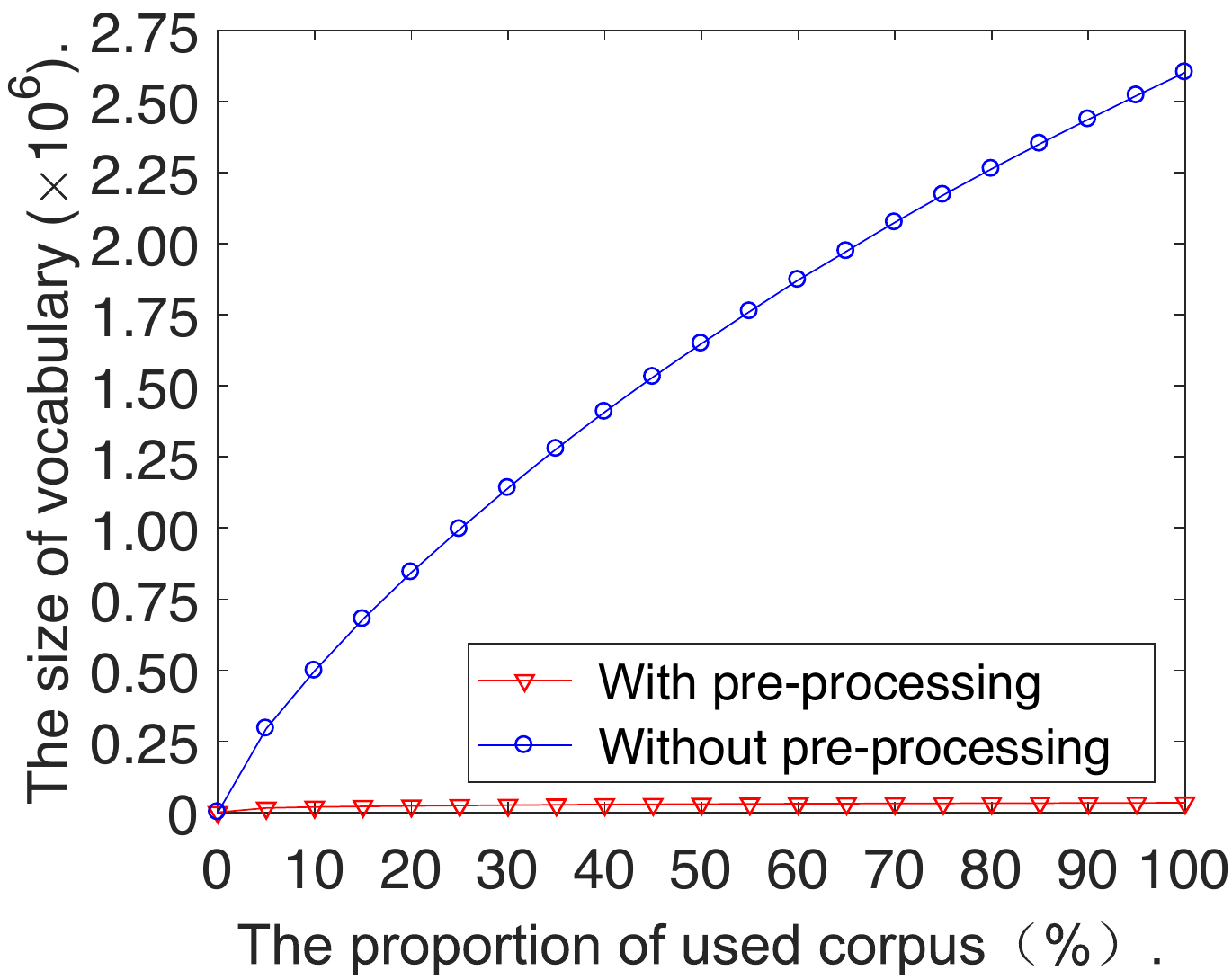}
\af
\caption{The growth of \zedit{the} vocabulary size.}\label{fig:voc_size}
\end{minipage}%
\\
\begin{minipage}{0.5\textwidth}
\centering
\includegraphics[scale=0.38]{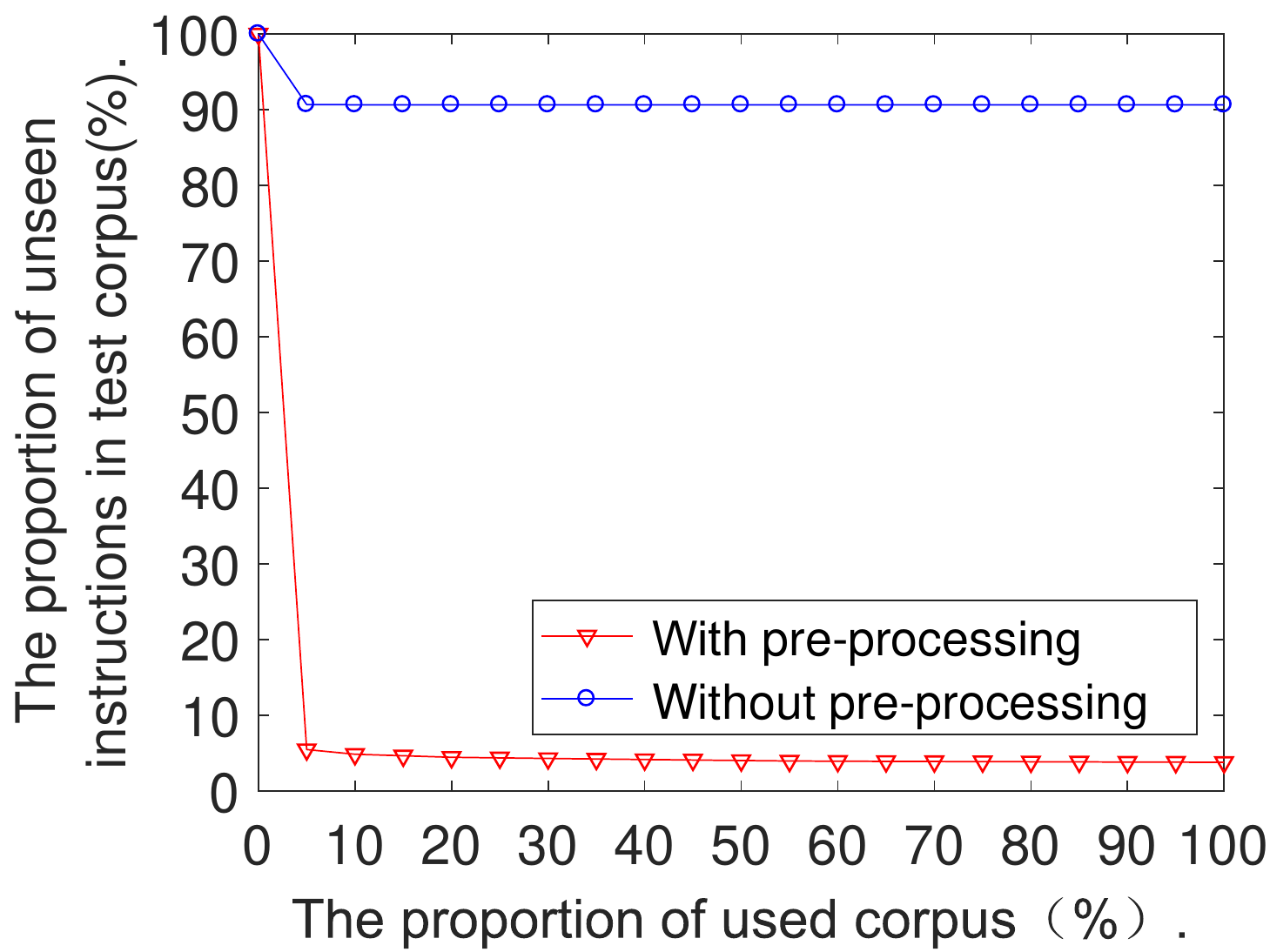}
\aaf
\caption{The proportion of unseen instructions.}\label{fig:voc_unseen}
\end{minipage}
\aaf\aaf\aaf
\end{figure}

To this end, we collect various x86 binaries, and disassemble these binaries to 
generate a corpus which contains 6,115,665 basic blocks and 39,067,830 assembly instructions.
We then divide the corpus equally into \zedit{20} parts. We counted the vocabulary size in terms of
the percentage of the corpus analyzed, and show the result in Figure~\ref{fig:voc_size}. The 
red line and the blue line show the growth of the vocabulary size when pre-processing is and
is not applied, respectively. It can be seen that the vocabulary size grows fast and becomes 
uncontrollable when the corpus is not pre-processed.

We next investigate the number of OOV cases, i.e., unseen instructions, in later
instruction embedding generation. We select two binaries that have never appeared in the 
previous corpus, \zedit{containing} 67,862 blocks and 453,724 instructions. 
We then count the percentage of unseen instructions that do not exist in the vocabulary, 
and show the result in Figure~\ref{fig:voc_unseen}. The red and blue lines show the 
percentage of unseen instructions when the vocabulary is built with or without pre-processing,
respectively. We can see that after pre-processing, only 3.7\% unseen instructions happen 
in later instruction embedding generation, \zedit{compared to 90\% without pre-processing}; 
(for an OOV instruction, a zero vector is assigned).
This shows that \zedit{the instruction embedding model with pre-processing} 
has a good coverage of instructions. Thus, it may be
reused by other researchers and we have made it publicly available.

\subsection{Qualitative Analysis of Instruction Embeddings}  \label{subsec:eval-instru-embed}

\begin{figure*}
    \centering
    \begin{minipage}{.6\textwidth}
	\includegraphics[height=4.2cm]{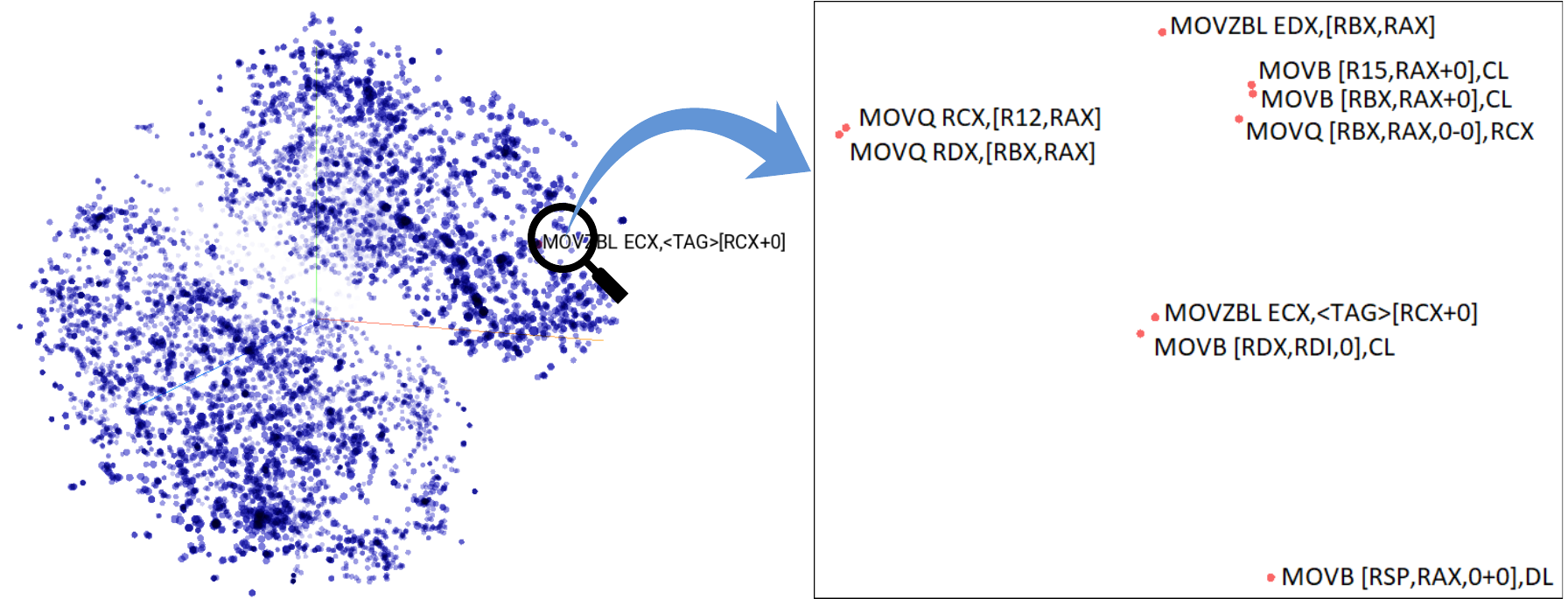} 
	\label{fig:w2v}	
	\caption{Visualization of all the instructions for x86 and ARM in 3D space, and a particular x86 instruction and its neighbor instructions, with t-SNE. \\ } \label{fig:w2v}
    \end{minipage}%
\quad
    \begin{minipage}{.35\textwidth}
	\centering
	\includegraphics[height=4.2cm]{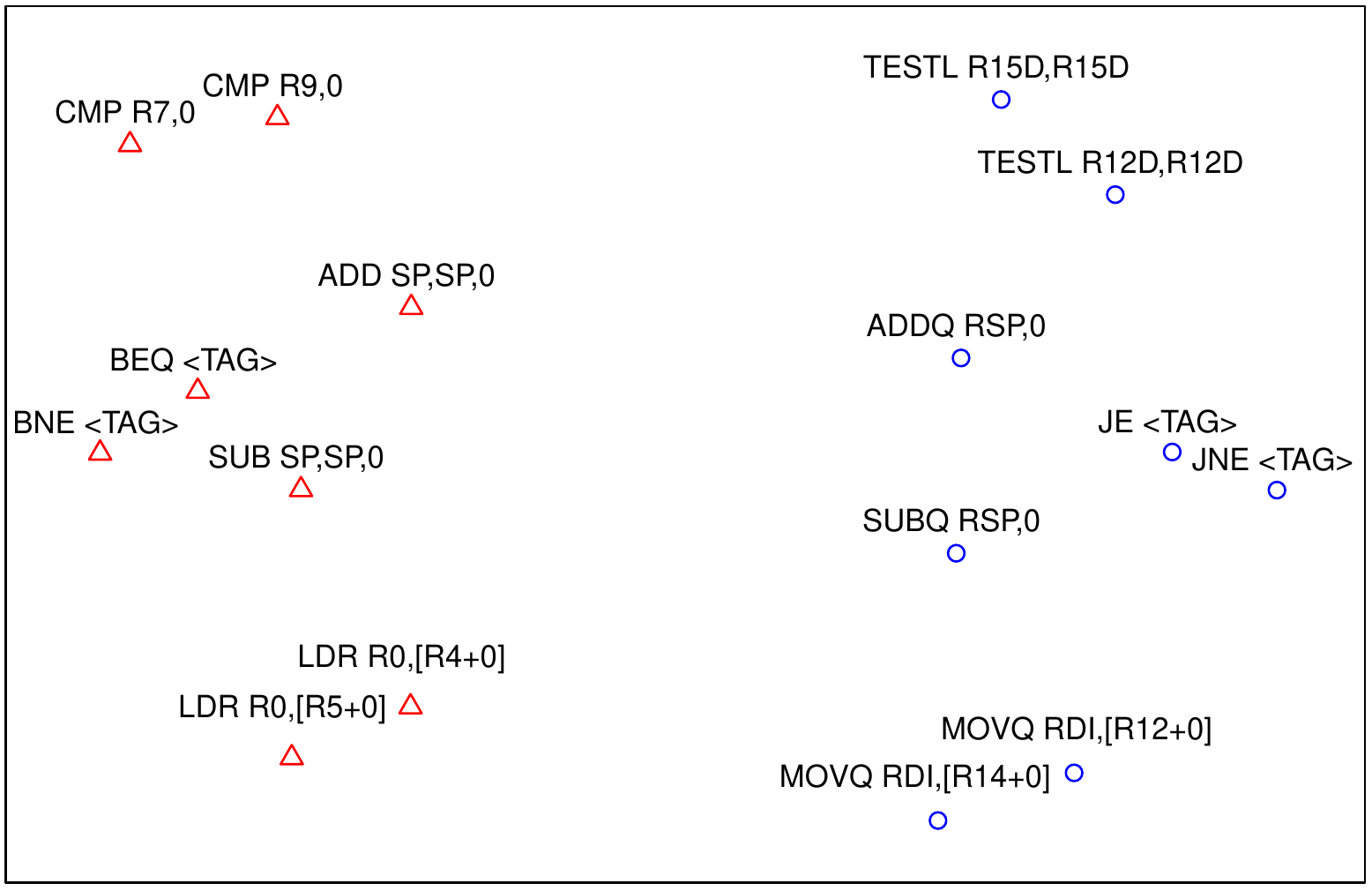}
	\caption{Visualization of a set of instructions for x86 and ARM based on MDS. The blue circles and red triangles 
represent x86 instructions and ARM instructions, respectively.}\label{fig:MDS}
    \end{minipage}%
\aaf\aaf\aaf
\end{figure*}

We present our results from qualitatively analyzing the instruction embeddings for the two 
architectures, x86 and ARM. We first use t-SNE~\cite{maaten2008visualizing}, a useful tool 
for visualizing high-dimensional vectors, to plot the instruction embeddings in a three-dimensional
space, as shown in Figure~\ref{fig:w2v}. A quick inspection immediately shows that the instructions 
compiled for the same architecture \zedit{cluster} together. Thus the most significant factor that 
influences code is the architecture as it introduces more syntactic variation. 
This also reveals \zedit{one of the reasons} why cross-architecture code similarity detection 
is more difficult than single-architecture code similarity detection. 

We then zoom in Figure~\ref{fig:w2v}, and plot a particular x86 instruction \texttt{MOVZBL EXC,<TAG>[RCX+0]}
and its neighbors. We can see that the \texttt{mov} 
family instructions are close together.  

\zedit{Next, we} {\color{black} use the analogical reasoning to evaluate the
quality of the cross-architecture instruction embedding model.} 
To do this, we randomly pick up eight x86 instructions. For each x86 instruction, we select its 
similar counterpart from ARM based on our prior knowledge and experience. 
We use [$x$] and \{$y$\} to represent the embedding of 
an ARM instruction $x$, and an x86 instruction $y$, respectively; and \texttt{cos}([$x_1$], [$x_2$]) 
refers to the cosine distance between two ARM instructions, $x_1$ and $x_2$. We have the following 
findings: (1) \texttt{cos}([\texttt{ADD SP,SP,0}], [\texttt{SUB SP,SP,0}]) is approximate to 
\texttt{cos}(\{\texttt{ADDQ RSP,0}\}, \{\texttt{SUBQ RSP,0}\}). 
(2) \texttt{cos}([\texttt{ADD SP,SP,0}], \{\texttt{ADDQ RSP,0}\}) is approximate to  
\texttt{cos}([\texttt{SUB SP,SP,0}], \{\texttt{SUBQ RSP,0}\}).
This is similar to other instruction pairs. {\color{black}We plot the relative positions of these instructions in Figure~\ref{fig:MDS} 
according to their cosine distance matrix based on MDS. 
} 
\todo{We limit the presented examples to eight due to space limitation. In our manual 
investigation, we find many such semantic analogies that are automatically learned.} 
Therefore, it shows that the instruction embedding 
model learns semantic information of instructions.

\subsection{Accuracy of \bbtool}  \label{subsec:eval-accuracy}

We now evaluate the accuracy of our \bbtool. All evaluations in this subsection are 
conducted on Dataset I.

\subsubsection{Model Training}  \label{subsec:eval-training}

We divide Dataset I into three parts for training, validation, and testing: for similar 
basic-block pairs, 80\% of them are used for training, 10\% for validation, and the remaining 
10\% for testing; the same splitting rule is applied to the dissimilar block pairs as well. 
Table~\ref{tab:dataset} shows the statistic results. In total, we have four training datasets: 
the first three datasets contain the basic-block pairs compiled with the same optimization level 
(O1, O2, and O3), and the last one contains the basic-block pairs with each one compiled with a different 
optimization level (cross-opt-levels). Note that in all the datasets, the two blocks of each 
pair are compiled for different architectures. This is the same for validation and testing 
datasets. 
Note that we make sure the training, validation, and testing datasets contain disjoint
sets of basic blocks (we split basic blocks into three disjoint sets before 
constructing similar/dissimilar
basic block pairs). Thus, any given basic block that appears in the training dataset
does not appear in the validation or testing dataset. Through this, we can 
better examine whether our model can work for unseen blocks. Note that the instruction embedding 
matrices for different architectures can be precomputed and reused.

We use the four training datasets to train \bbtool\ individually for 100 epochs. After each 
epoch, we measure the AUC and loss on the corresponding validation datasets, and save the 
models achieving the best AUC as the base models.

\subsubsection{Results} \label{subsec:eval-result}

\begin{figure*}
\centering
\subfloat[O1]{\includegraphics[width=1.85in]{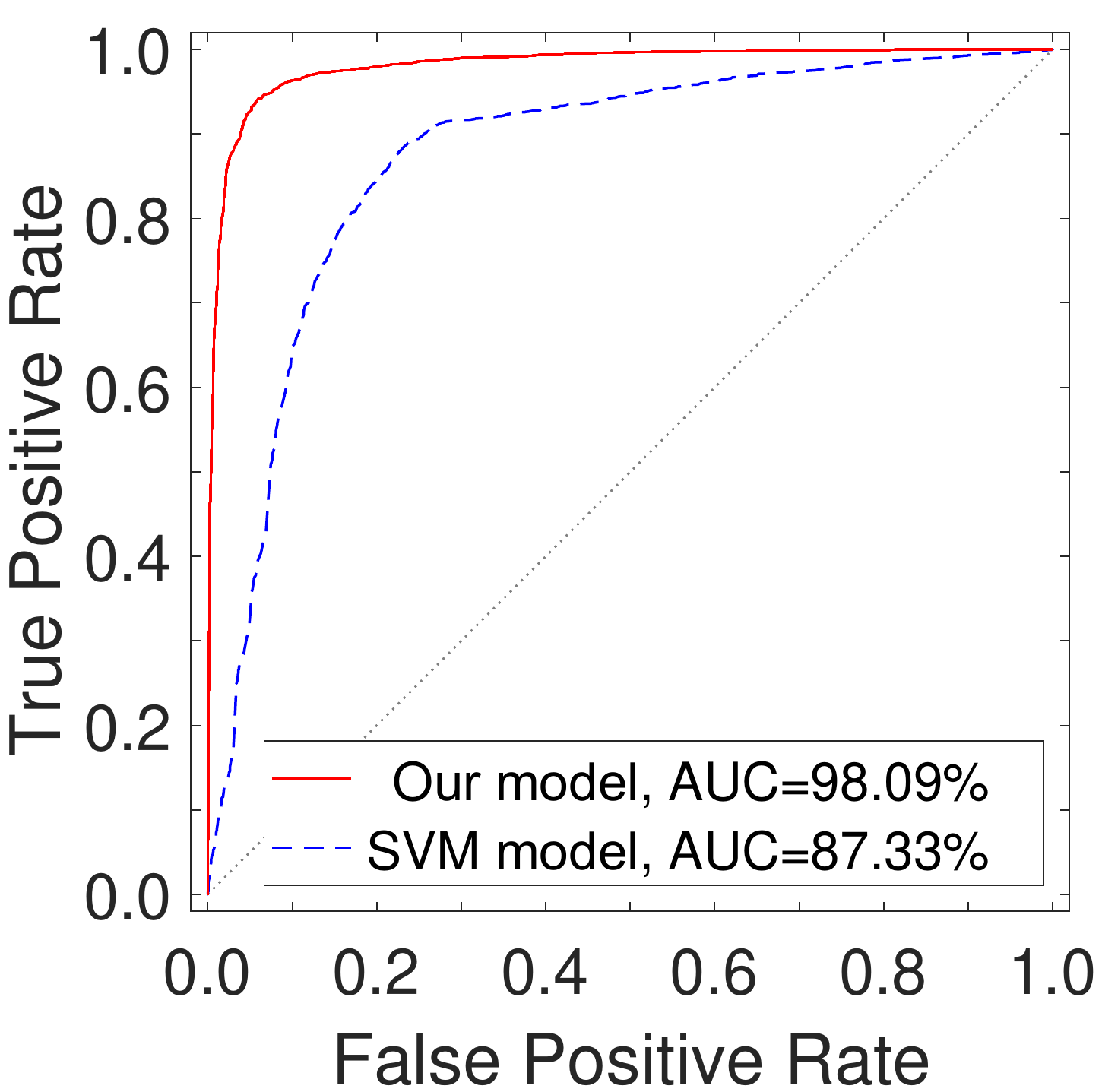}}\qquad
\subfloat[O2]{\includegraphics[width=1.85in]{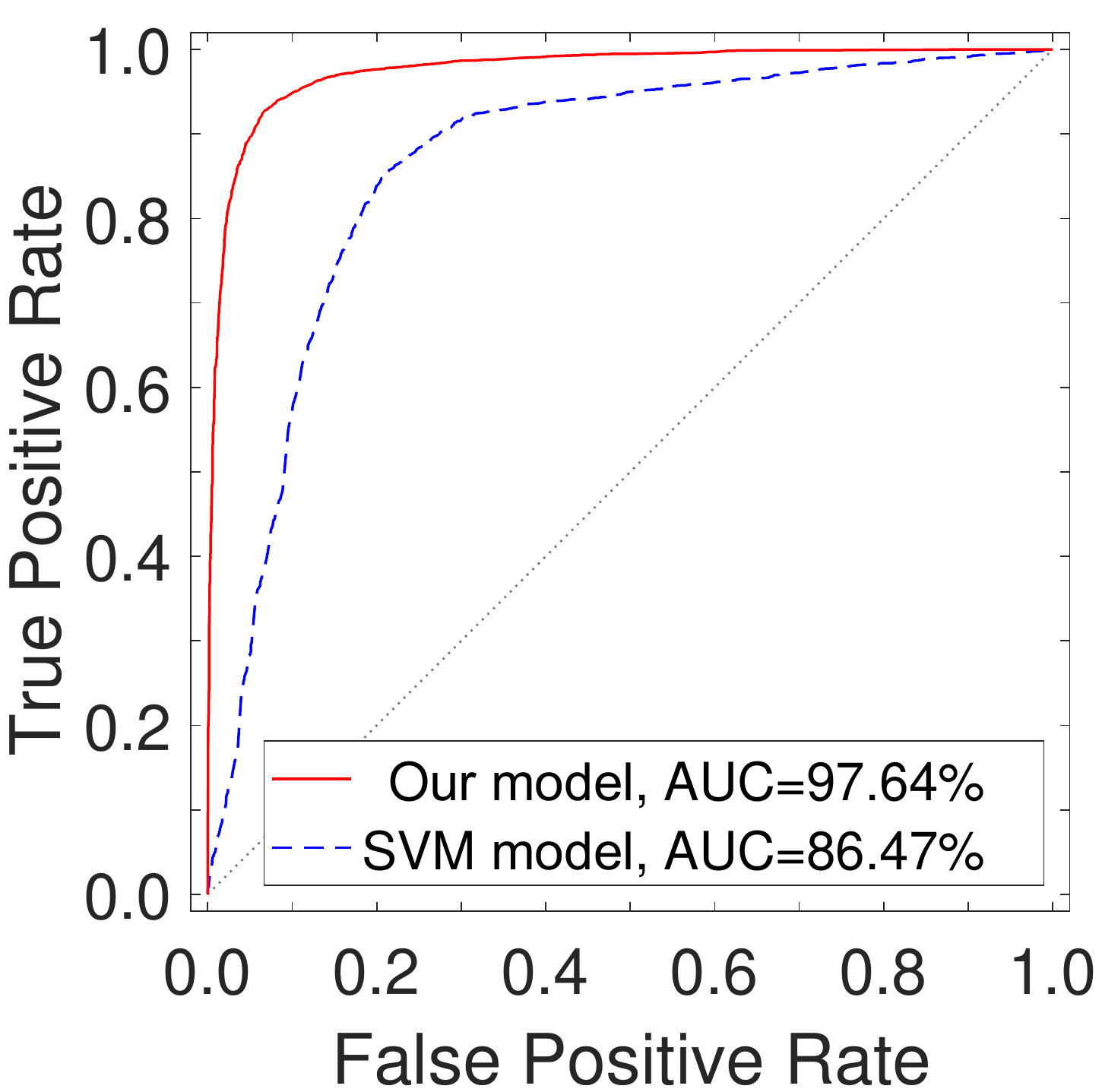}}\qquad
\subfloat[O3]{\includegraphics[width=1.85in]{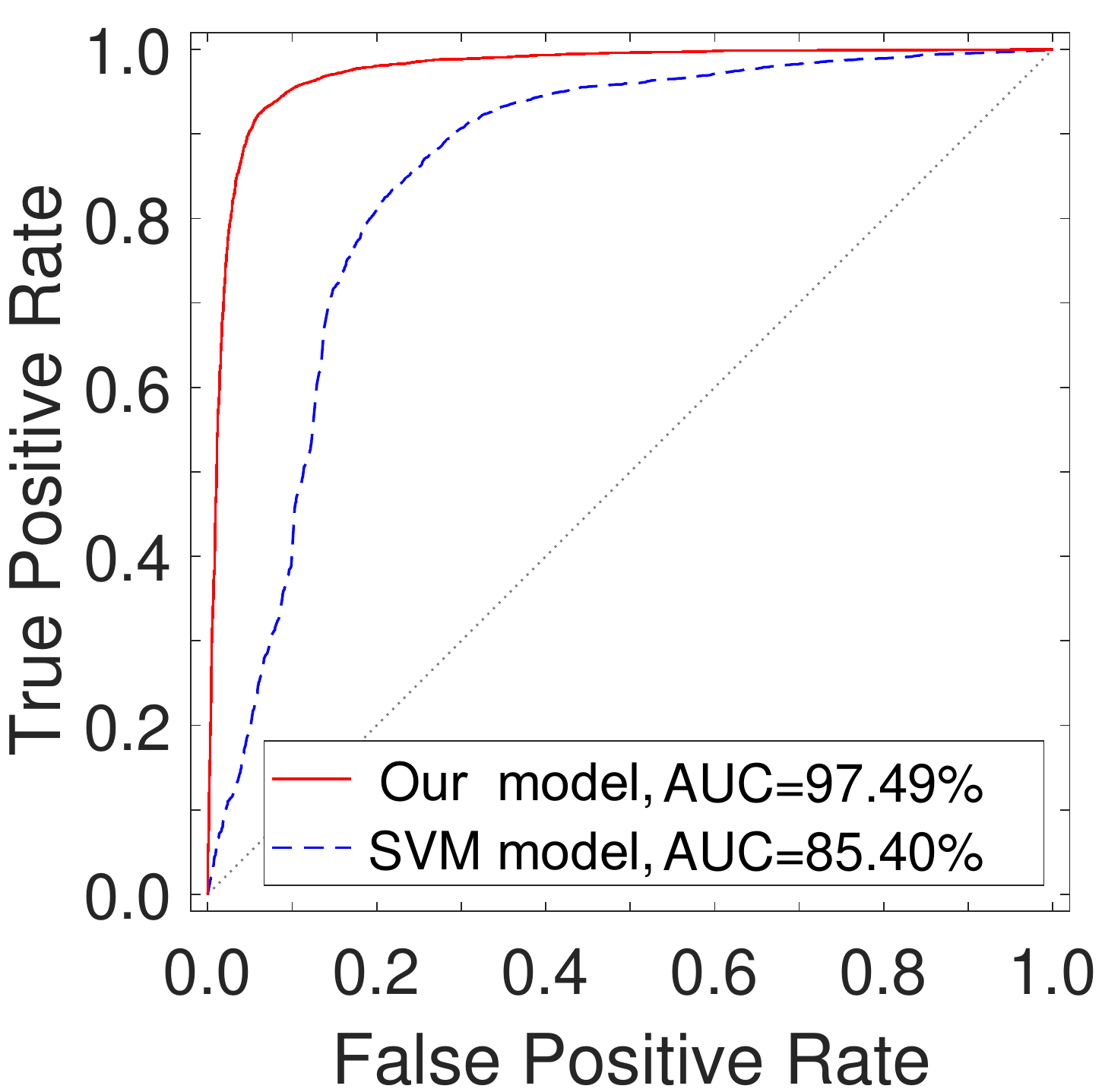}}
\\
\af
\subfloat[Cross-opt-levels]{\includegraphics[width=1.85in]{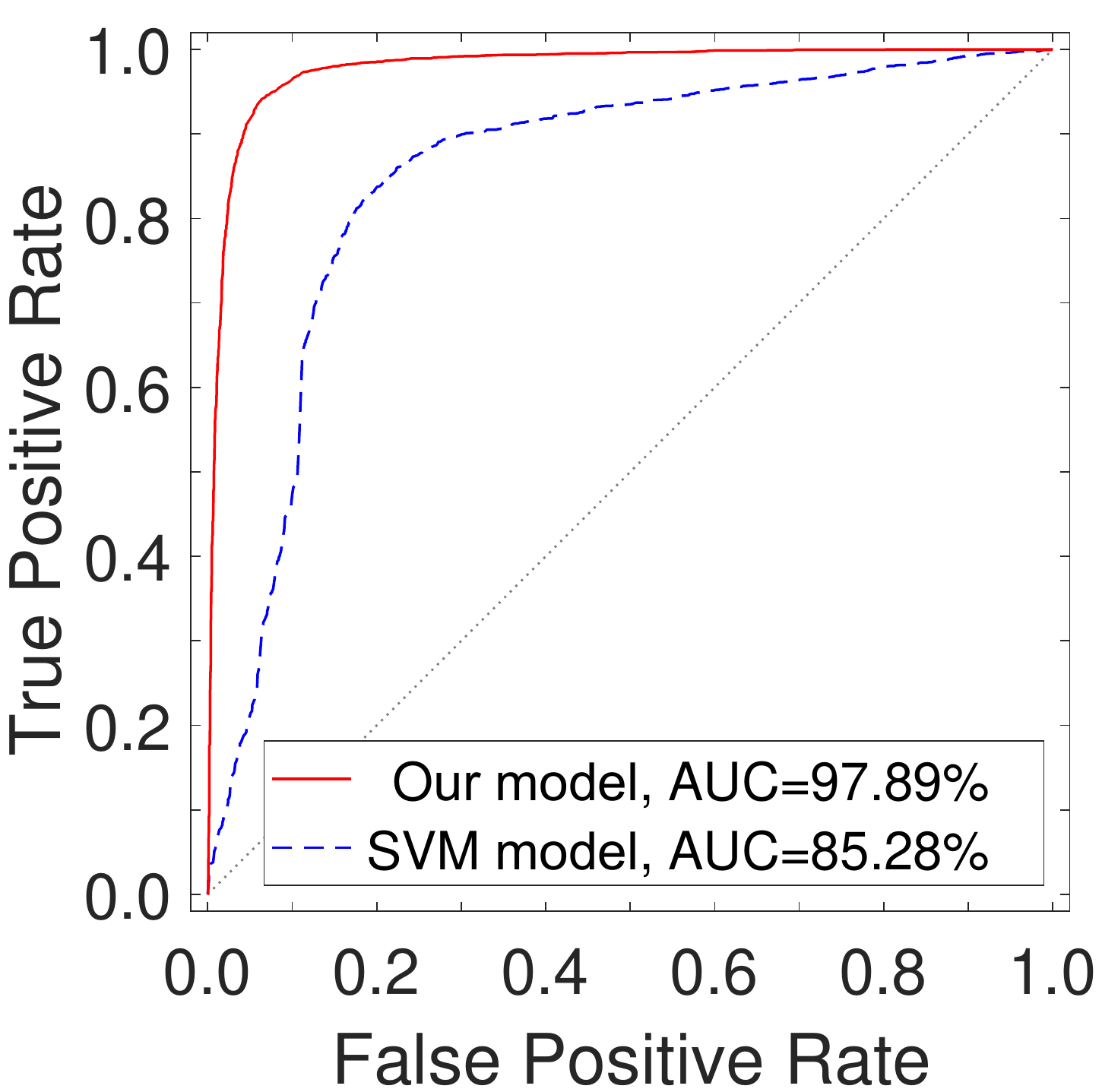}} \qquad
\subfloat[{\color{black}Large basic blocks in O3}]{\includegraphics[width=1.85in]{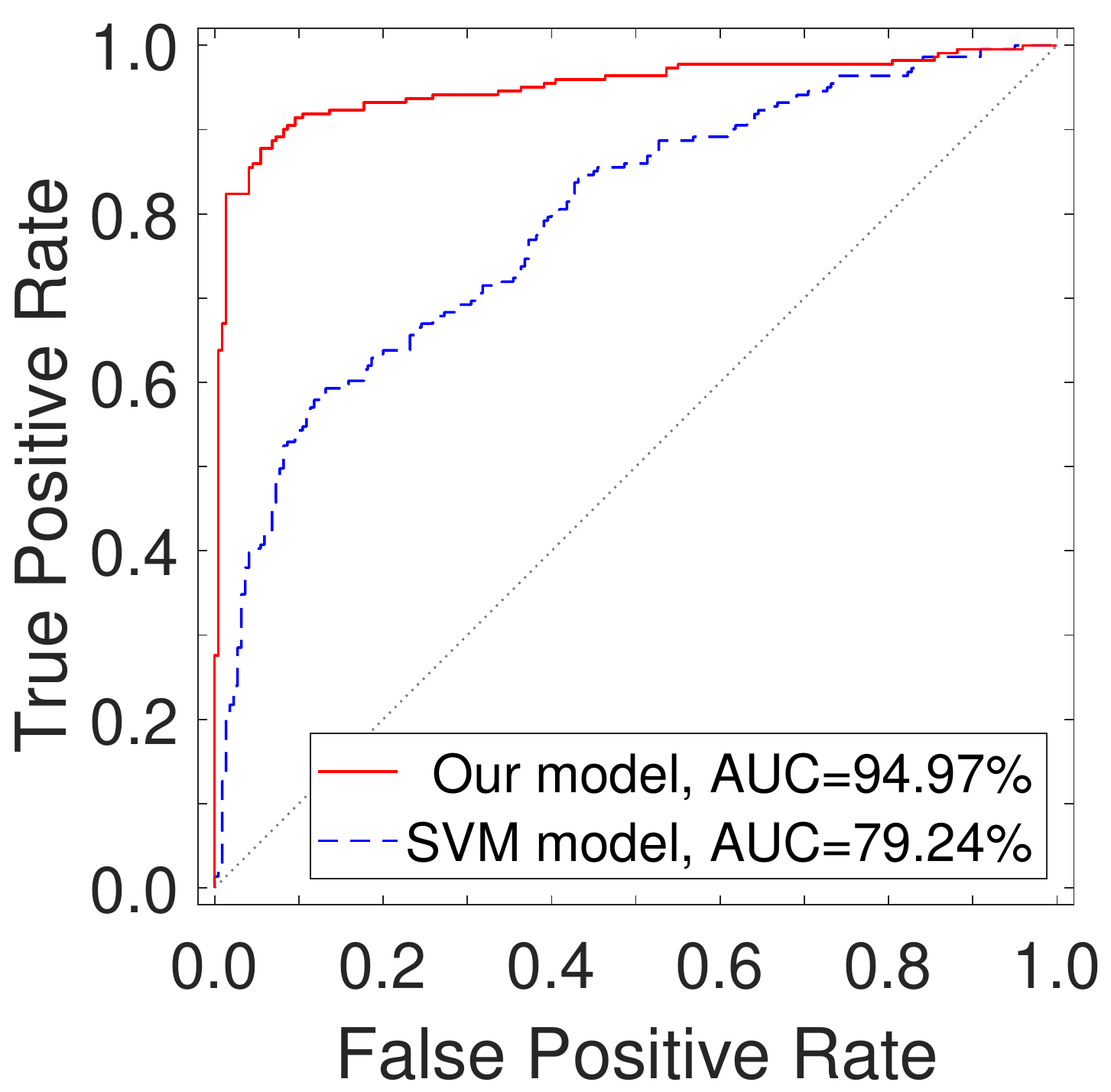}}\qquad
\subfloat[{\color{black}Small basic blocks in O3}]{\includegraphics[width=1.85in]{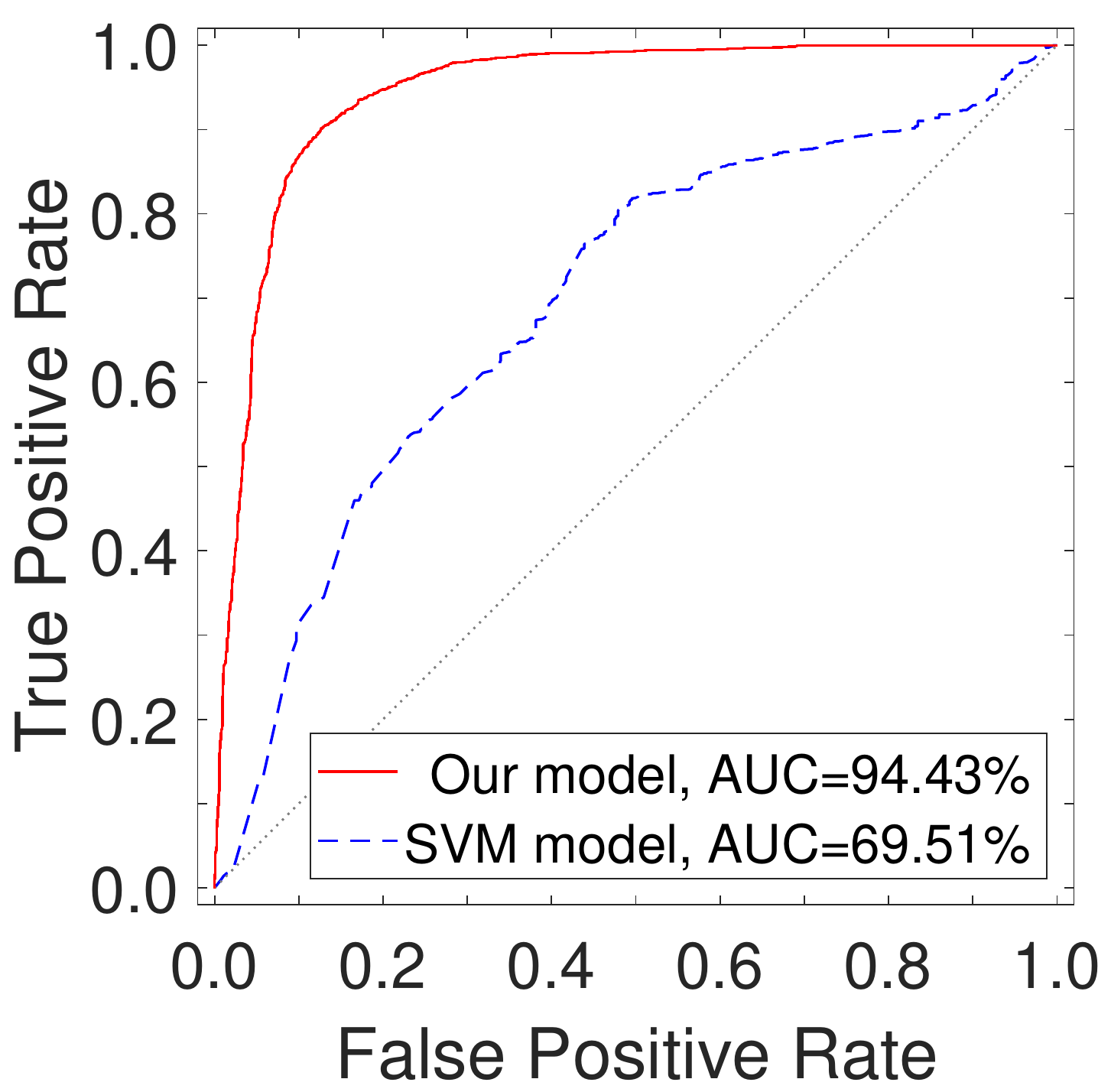}} 
\caption{The ROC evaluation results based on the four testing datasets.} \label{fig:ROC}
\aaf\aaf
\end{figure*}

We now evaluate the accuracy of the base models using the corresponding testing datasets. 
The red lines in the first four figures in Figure~\ref{fig:ROC}, from (a) to (d), are the 
ROC {\color{black}curves} of the similarity test. As each curve is close to the left-hand and top border, 
our models have good accuracy.

To further comprehend the performance of our models on basic blocks with different sizes, we create
 \zedit{small-BB and large-BB} testing subsets. 
 If a basic block contains less than 5 instructions it 
belongs to the small-BB subset; a block containing more than 20 instructions belongs to the 
large-BB subset. We then evaluate the corresponding ROC. Figure~\ref{fig:ROC}e and Figure~\ref{fig:ROC}f 
show the ROC results evaluated on the large-BB subset (221 pairs) and small-BB subset (2409 
pairs), respectively, where the basic-block pairs are compiled with the O3 optimization level. 
The ROC results evaluated on the basic-block pairs compiled with other optimization levels are 
similar, and are omitted here due to \zedit{the page limit}. We can observe that our models achieve 
good accuracy for both small blocks and large ones. 
\ledit{Because a small basic block contains less semantic information, the AUC (=94.43\%) of the small-BB
subset (Figure~\ref{fig:ROC}f) is slightly lower than others. Moreover, as there are a small
portion (4.4\%) of large BB pairs in the training dataset, the AUC (=94.97\%) of the large-BB 
subset (Figure~\ref{fig:ROC}e) is also slightly lower; we expect this could be improved if more large
BB pairs are seen during training.}

\subsubsection{Comparison with Manually Selected Features} \label{subsec:eval-comparison}

Several methods are proposed for cross-architecture basic block similarity detection, e.g.,
fuzzing~\cite{pewny2015cross}, symbolic execution~\cite{gao2008binhunt,luo2014semantics}, and 
basic-block feature-based machine learning classifier~\cite{feng2016scalable}. Fuzzing and symbolic 
execution are much slower than our deep learning based approach. We thus compare our model against the SVM classifier using six 
\emph{manually} selected block features adopted in \texttt{Gemini}, such as the number of instructions and the number of constants (see Table 1 in ~\cite{ccs2017graphembedding}. 
 
We extract the six features from 
each block to represent the block, and use all blocks in the training dataset to train the SVM classifier. 
We adopt leave-one-out cross-validation with $K=5$ and use the Euclidean distance to measure 
the similarity of blocks. By setting the complexity parameter 
$c=1.0$, $\gamma=1.0$ and choosing the \texttt{RBF} kernel, the SVM classifier achieves the best 
AUC value. Figure~\ref{fig:ROC} shows the comparison results on different testing subsets. 
We can see that our models outperform the SVM classifier and achieve much higher 
AUC values. This is because the \emph{manually} selected features \zedit{largely lose 
the instruction semantics and dependency information}, while \bbtool\ precisely
encodes the block semantics.

\noindent \textbf{Examples.}
\ledit{Table~\ref{table:block_examples-similar} shows three pairs of \emph{similar} basic-block pairs 
(after pre-processing) that are correctly classified by \bbtool, but misclassified by the 
statistical feature-based SVM model. Note that the pre-processing does not change the statistical
features of basic blocks; e.g., the number of transfer instructions keeps the same before
and after pre-processing. 
Our model correctly reports each pair as similar.}

\ledit{Table~\ref{table:block_examples-dissimilar} shows three pairs of \emph{dissimilar} basic-block 
pairs (after pre-processing) that are correctly classified by \bbtool, but misclassified by the 
SVM model. As the statistical features of two dissimilar blocks 
tend to be similar, the SVM model---which ignores the meaning of instructions and the dependency 
between them---misclassifies them as similar. }

\begin{table*}[t]
    \centering
    \begin{minipage}{1\textwidth}
\footnotesize
\caption{Examples of \emph{similar} BB pairs that are correctly classified by \bbtool, but misclassified by the SVM model.}\label{table:block_examples-similar}
\centering
\renewcommand{\arraystretch}{1.07}
\scalebox{1}{
\begin{tabular}{@{}l@{}|@{~}l@{}|@{~}l@{}|@{~}l@{}|@{~}l@{}|@{~}l@{}}
\hline
\multicolumn{2}{c|}{\textbf{Pair 1}} & \multicolumn{2}{c|}{\textbf{Pair 2}}  & \multicolumn{2}{c}{\textbf{Pair 3}} \\  \hline
\multicolumn{1}{c|}{\emph{x86}}  & \multicolumn{1}{c|}{\emph{ARM}}  & \multicolumn{1}{c|}{\emph{x86}} & \multicolumn{1}{c|}{\emph{ARM}} & \multicolumn{1}{c|}{\emph{x86}} & \multicolumn{1}{c}{\emph{ARM}}   \\
\hhline{=|=|=|=|=|=}
MOVSLQ RSI,EBP       & LDRB R0,[R8+R4]  & MOVQ RDX,<TAG>+[RIP+0] & LDR R2,[R8+0]  & MOVQ [RSP+0],RBX  & LDR R0,[SP+0]       \\
MOVZBL ECX,[R14,RBX] & STR R9,[SP]      & MOVQ RDI,R12           & MOV R0,R4      & MOVQ [RSP+0],R14  & STR R9,[SP+0]    \\
MOVL EDX,<STR>       & STR R0,[SP+0]    & MOVL ESI,R14D          & MOV R1,R5      & ADDQ RDI,0        & STR R0,[SP+0]    \\
XORL EAX,EAX         & ASR R3,R7,0      & CALLQ FOO              & BL FOO         & CALLQ FOO         & ADD R0,R1,0       \\
MOVQ RDI,R13         & MOV R0,R6        & MOVQ RDI,R12           & MOV R0,R4      & MOVL ESI,<TAG>    & BL FOO           \\
CALLQ FOO            & MOV R2,R7        & CALLQ FOO              & BL FOO         & MOVQ RDI,[R12]    & LDR R7,<TAG>      \\
TESTL EAX,EAX        & BL FOO           & MOVQ RDX,<TAG>+[RIP+0] & LDR R2,[R8+0]  & MOVB [RDI+0],AL   & LDR R1,[R6]        \\
JLE <TAG>            & CMP R0,0         & MOVQ RDI,R12           & MOV R0,R4      & CMPB [RDI+0],0    & LDR LR,[SP+0]     \\
                     & BLT <TAG>        & MOVL ESI,R14D          & MOV R1,R5      & JNE <TAG>         & MOV R12,R7         \\
                     &                  & CALLQ FOO              & BL FOO         &                   & STRB R0,[R1+0]  \\
                     &                  & TESTL EAX,EAX          & CMP R0,0       &                   & B <TAG>           \\
                     &                  & JNE <TAG>              & BNE<TAG>       &                   &                 \\
\hline
\end{tabular}
}
\vspace{3pt}
\end{minipage}
~
\begin{minipage}{1\textwidth}
\footnotesize
\caption{Examples of \emph{dissimilar} BB pairs that are correctly classified by \bbtool, but misclassified by the SVM model.}\label{table:block_examples-dissimilar}
\centering
\renewcommand{\arraystretch}{1.07}
\scalebox{1}{
\begin{tabular}{@{}l@{}|@{~}l@{}|@{~}l@{}|@{~}l@{}|@{~}l@{}|@{~}l@{}}
\hline
\multicolumn{2}{c|}{\textbf{Pair 4}} & \multicolumn{2}{c|}{\textbf{Pair 5}}  & \multicolumn{2}{c}{\textbf{Pair 6}} \\ \hline
\multicolumn{1}{c|}{\emph{x86}}  & \multicolumn{1}{c|}{\emph{ARM}}  & \multicolumn{1}{c|}{\emph{x86}} & \multicolumn{1}{c|}{\emph{ARM}} & \multicolumn{1}{c|}{\emph{x86}} & \multicolumn{1}{c}{\emph{ARM}}   \\
\hhline{=|=|=|=|=|=}
       IMULQ R13,RAX,0   & MOV R1,R0      & XORL R14D,R14D   & LDMIB R5,{R0,R1} &  MOVL [RSP+0],R14D    & SUB R2,R1,0 \\
       XORL EDX,EDX      & LDR R6,[SP+0]  & TESTQ RBP,RBP    & CMP~R0,R1        &  MOVQ RAX,[RSP+0]     & MOV R10,0 \\
       MOVQ RBP,[RSP+0]  & CMP R0, 0      & JE <TAG>         & BHS <TAG>        &  CMPB [RAX],0         & CMP R2,0 \\
       DIVQ RBP          & BEQ <TAG>      &                  &                  &  MOVQ [RSP+0],R13     & MOV R9,0 \\
       JMP <TAG>         &                &                  &                  &  MOVQ [RSP+0],R15     & BHI <TAG> \\
                         &                &                  &                  &  JNE <TAG>            &  \\
\hline
\end{tabular}
}
\vspace{3pt}
\end{minipage}
\end{table*}

\subsection{Hyperparameter Selection for \bbtool}  \label{subsec:eval-hyper}

We next investigate the impact of different hyperparameters on \bbtool. 
In particular, we consider the number of epochs, the dimensionality 
of the embeddings, network depth, and hidden unit types. 
We use the validation datasets of Dataset I to examine the impact of the number of epochs,
and the testing datasets to examine the impact of other hyperparameters.

\begin{figure*}
\subfloat[AUC vs. \# of epochs.]{%
	\adjustbox{raise=-3.8pc}{\includegraphics[scale=0.41]{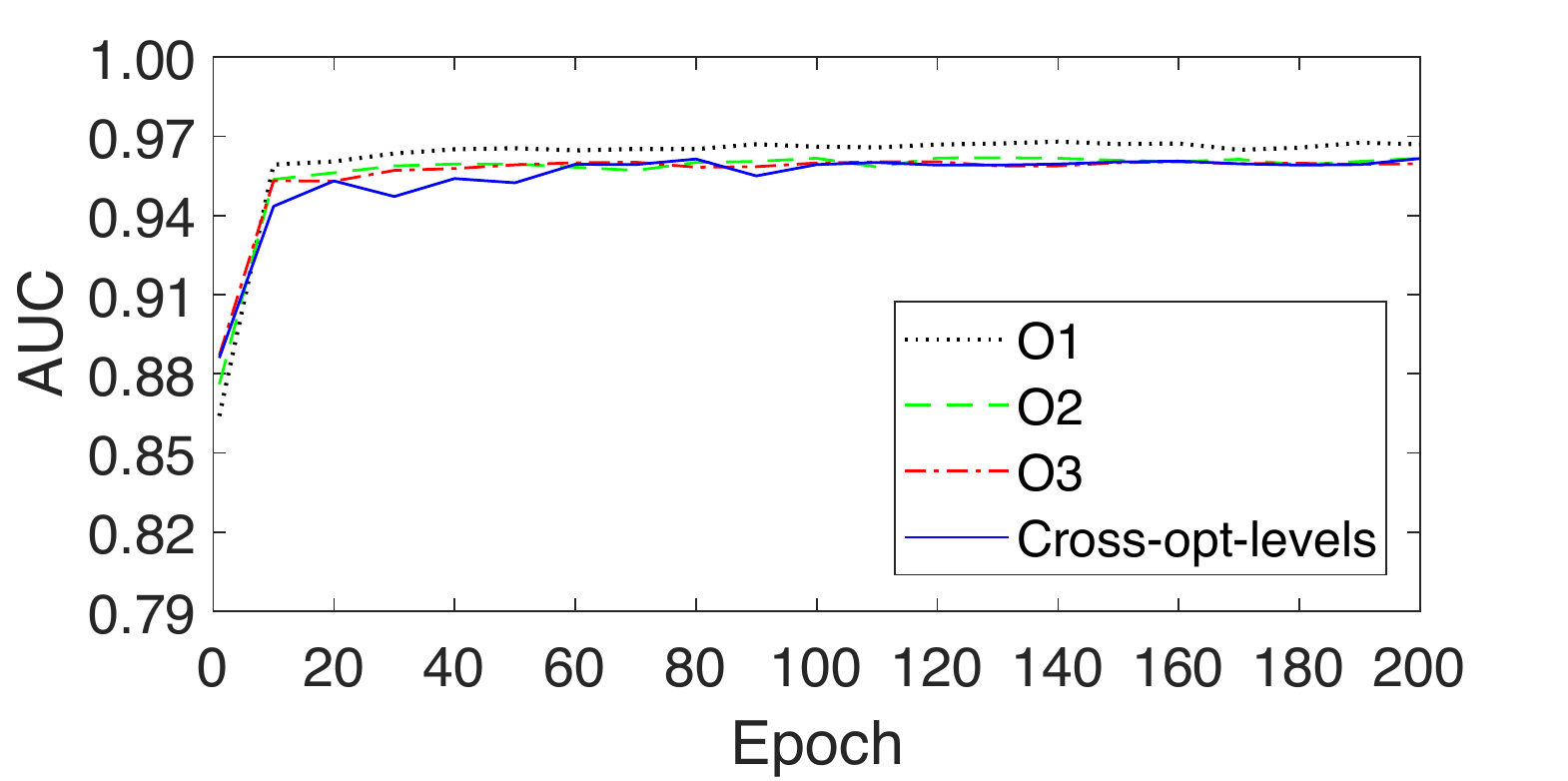}}%
	\label{subfig-1}
	}
~	
\subfloat[Loss vs. \# of epochs.]{%
	\adjustbox{raise=-3.8pc}{\includegraphics[scale=0.41]{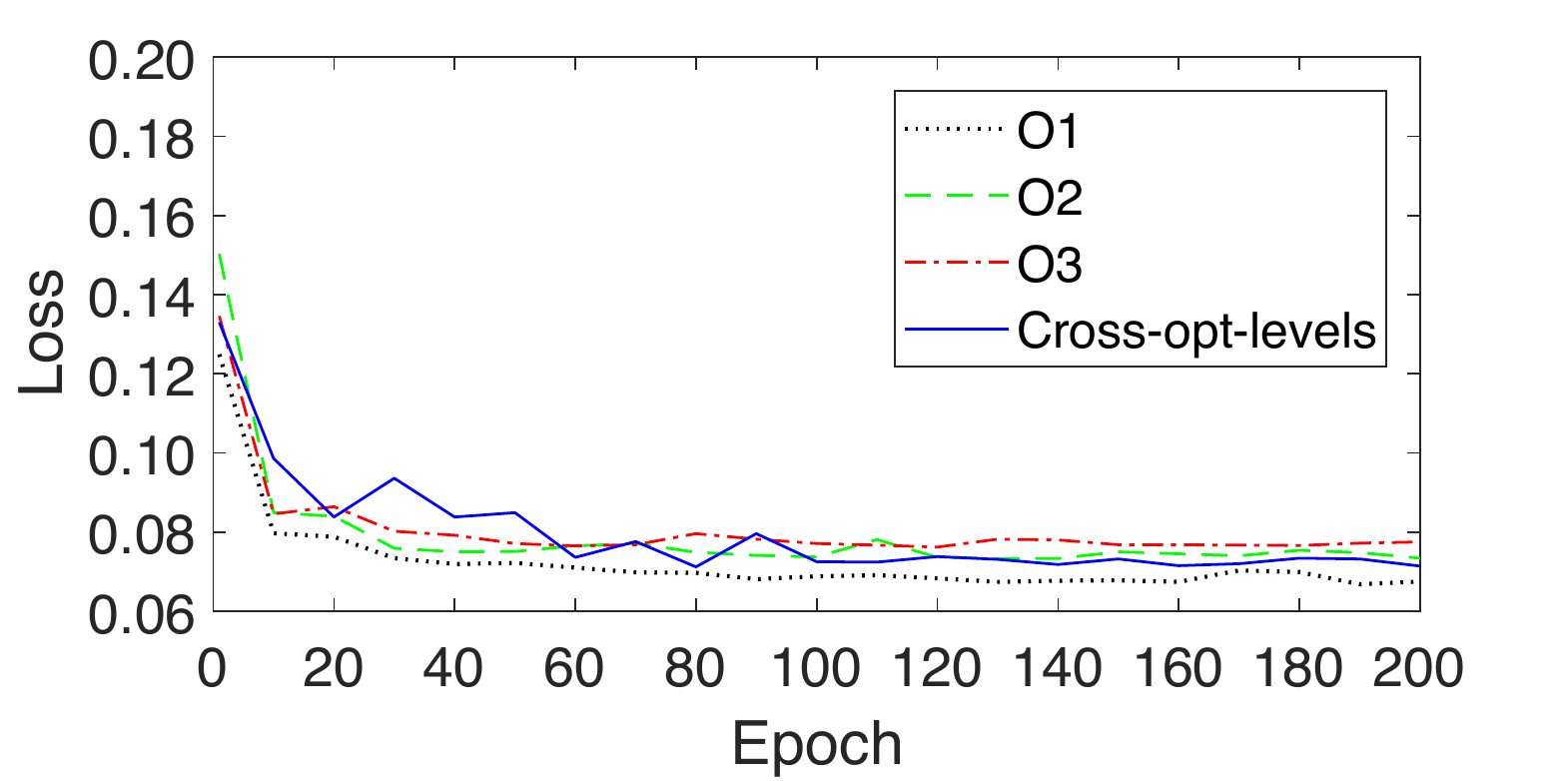}}%
	\label{subfig-2}%
	}
~ 
\subfloat[AUC vs. instruction embedding dimensions.]{%
	\small \scalebox{1}{
	\begin{tabular}{|c|c|c|c|c|}
	\hline
	\multirow{2}*{ (\%) }  & \multicolumn{3}{c|}{Optimization levels} & Cross\\
	\cline{2-4}
	\multirow{3}*{~}   & \multirow{1}*{O1}      & \multirow{1}*{O2}      & \multirow{1}*{O3}      & -opts \\ 
	\hline
	50 & 95.77 & 95.23 & 94.97 & 95.39  \\ 
	100 & 96.83 & 96.33 & 95.99 & 95.82  \\
	150 & 96.89 & 96.33 & 96.24 & 95.86  \\ 
	\hline
	\end{tabular}}
	\vspace{6pt}
	\label{tab: inputsize}%
	}

\subfloat[AUC vs. block embedding dimensions.]{%
	\small \scalebox{1}{
	\begin{tabular}{|c|c|c|c|c|}
	\hline
	\multirow{2}*{ (\%) }  & \multicolumn{3}{c|}{Optimization levels} & Cross\\
	\cline{2-4}
	\multirow{2}*{~}        & O1      & O2      & O3      &  -opts \\ \hline
	10 & 95.57 & 95.73 & 95.48 & 95.59  \\ 
	30 & 95.88 & 95.65 & 96.17 & 95.45  \\ 
	50 & 96.83 & 96.33 & 95.99 & 95.82  \\    
	\hline
	\end{tabular}}
	\label{table_HU}%
	}
~
\subfloat[AUC vs. \# of network layers.]{%
	\small \scalebox{1}{
	\begin{tabular}{|c|c|c|c|c|}
	\hline
	\multirow{2}*{ (\%) }  & \multicolumn{3}{c|}{Optimization levels} & Cross\\
	\cline{2-4}
	\multirow{2}*{~}        & O1      & O2      & O3      & -opts \\ \hline
	1 & 95.88 & 95.65 & 96.17 & 95.45  \\ 
	2 & 97.83 &  97.49 &  97.59 &  97.45  \\     
	3 & 98.16 & 97.39 & 97.48 & 97.76  \\ 
	\hline
	\end{tabular}}
	\label{table_HL}%
	}
~
\subfloat[AUC vs. network hidden unite types.]{%
	\small \scalebox{1}{
	\begin{tabular}{|c|c|c|c|c|}
	\hline
	\multirow{2}*{ (\%) }  & \multicolumn{3}{c|}{Optimization levels} & Cross\\
	\cline{2-4}
	\multirow{2}*{~}        & O1      & O2      & O3      &  -opts \\ \hline
	LSTM & 96.83 & 96.33 & 95.99 & 95.82  \\ 
	GRU & 96.15 & 95.30 & 95.83 & 95.71  \\ 
	RNN & 91.39 & 93.26 & 92.60 & 92.66  \\ 
	\hline
	\end{tabular}}
	\label{table_UTypes}%
	}
\aaf
\vspace{9pt}
\caption{Impact of different hyperparameters. Figure~\ref{subfig-1} and
Figure~\ref{subfig-2} are evaluated on the validation datasets of Dataset I, and others are evaluated 
on its testing datasets.} \label{fig:epoch}
\aaf\af
\end{figure*}

\subsubsection{Number of Epochs}  \label{subsec:epoch}

To see whether the accuracy of the model fluctuates during training, we trained
the model for 200 epochs and evaluated the model every 10 epochs for the AUC and loss. 
The results are displayed in Figure~\ref{subfig-1} and Figure~\ref{subfig-2}.
We observe that the AUC value steadily increases and is stabilized 
at the end of epoch 20; and the loss value decreases quickly and almost stays stable 
after 20 epochs. Therefore, we conclude that the model can be quickly trained 
to achieve good performance.

\subsubsection{Embedding Dimensions} 

We next measure the impact of the instruction embedding and block embedding dimensions.

\noindent
\textbf{Instruction embedding dimension.} 
We vary the instruction embedding dimension, and evaluate the corresponding AUC values 
shown in Figure~\ref{tab: inputsize}. We observe that increasing the embedding dimensions 
yields higher performance; and the AUC values corresponding to the embedding dimension 
higher than 100 are close to each other. Since a higher embedding dimension leads to higher 
computational costs (requiring longer training time), we conclude that a moderate dimension 
of 100 is a good trade-off between \zedit{precision} and efficiency. 

\noindent
\textbf{Block embedding dimension.} 
Next, we vary the block embedding dimension, and evaluate the corresponding AUC values shown 
in Figure~\ref{table_HU}. We observe that the performance of the models with 10, 30 and 50 block 
embedding dimensions are close to each other. Since a higher embedding dimension leads to higher 
computational costs, we conclude that a dimension of 50 for block embeddings is a good trade-off. 

\subsubsection{Network Depth} \label{subsec:depth} 

We then change the number of layers of each {\color{black}LSTM}, and evaluate the corresponding AUC values.
Figure~\ref{table_HL} shows that the LSTM networks with two and three layers outperform the 
network with a single layer, and the AUC values for the networks with two and three layers 
are close to each other. Because adding more layers increases the computational complexity 
and does not help significantly on the performance, we choose the network depth as 2.

\subsubsection{Network Hidden Unit Types} 

{\color{black}As a simpler-version of LSTM, Gated Recurrent Unit (GRU) has become increasingly 
popular. We conduct experiments on comparing three types of network 
units, including LSTM, GRU as well as RNN. } Figure~\ref{table_UTypes} shows the comparison results. 
It can be seen that LSTM and GRU are 
more powerful than the basic RNN, and LSTM shows the highest AUC values. 

\subsection{Efficiency of \bbtool} \label{subsec:eval-efficiency}

\subsubsection{Training Time} 

We first analyze the training time for both the instruction and basic-block embedding models.

\vspace{3pt}
\noindent
\textbf{Instruction embedding model training time.}
The training time is linear to the number of epochs and the corpus size. We use Dataset I, 
containing 437,104 blocks for x86 and 393,529 blocks for ARM, with 6,199,651 instructions in total, 
as the corpus to train the instruction embedding model. The corpus contains 49,760 distinct 
instructions which form a vocabulary. We use $10^{-5}$ as the down sampling rate and set the 
parameter \texttt{mini-word-count} as zero (no word is ignored during training), and train
the model for 100 epochs. Table~\ref{tab: ins_emb_train_time} shows the training time with 
respect to different instruction embedding dimensions. We can see that the instruction embedding 
model can be trained in a very short period of time.

\begin{figure*}
    \centering
    \begin{minipage}{.61\textwidth}
        \subfloat[Training time of single-layer networks with respect to different hidden unit types.]{%
	\includegraphics[scale=0.43]{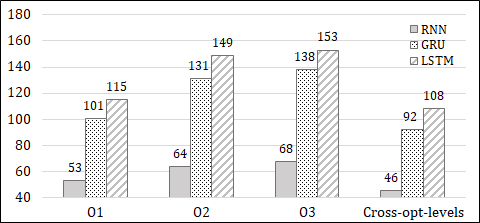}
	\label{subfig:time-1}
	}
	~
	\subfloat[Training time of LSTMs with respect to different number of network layers]{%
	\includegraphics[scale=0.43]{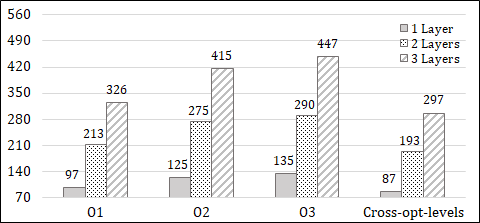}
	\label{subfig:time-2}	
	} 
	\caption{Training time of the basic-block embedding model. The instruction embedding dimension is 100, and the block embedding dimension is 50. The number above each bar is the time (\emph{second per epoch}) used to train the model.} \label{fig:train_time}
    \end{minipage}%
\quad
    \begin{minipage}{0.36\textwidth}
	\small \vspace{11pt}
        \begin{tabular}{|c|c|c|c|c|}
	\hline
	\multirow{2}*{ (\emph{Second}) }  & \multicolumn{3}{c|}{Optimization levels} & Cross\\
	\cline{2-4}
	\multirow{2}*{~}        & O1      & O2      & O3      & -opts \\ \hline
	L=1, D=30&3.040&3.899&  4.137 &  2.944 \\ 
	L=1, D=50 & 3.530 & 4.702 & 4.901 & 3.487  \\  
	L=2, D=30&6.359&8.237&  8.780& 6.266 \\  
	L=2, D=50 & 6.663 & 8.722 & 9.139 & 6.625  \\    
	\hline
	\end{tabular} 
	\vspace{1pt}
        \caption{Testing time of \bbtool\ with respect to different number of network layers and 
block embedding dimensions. The instruction embedding dimension is 100. $L$ denotes the number of network layers. $D$ 
denotes the block embedding dimension.}
        \label{tab:TestTime}
    \end{minipage}
\aaf\aaf
\end{figure*}

\vspace{3pt}
\noindent
\textbf{Block embedding model training time.} 
We next evaluate the time used for training the basic-block embedding model. The training time is linear 
to the number of epochs and the number of training samples. The results are showed in Figure~\ref{fig:train_time}. 
The number above each bar is the time (\emph{second per epoch}) used to train the model. Figure~\ref{subfig:time-1} 
shows the training time with respect to different types of network hidden unit. Figure~\ref{subfig:time-2} displays 
the training time of the LSTM networks in terms of different number of network layers. In general, LSTM takes 
longer training time, and a more complicated model (with more layers) requires more time per epoch. 

\begin{table}[!t]
\af
\caption{Training time of the instruction embedding model with respect to different embedding dimensions.}
\label{tab: ins_emb_train_time}
\af
\renewcommand{\arraystretch}{1.07}
\begin{tabular}{|c|ccc|}
\hline
Instruction embedding dimension & 50 &100 & 150\\
\hline
Training time (\emph{second}) & 82.71 & 84.22 & 89.75\\
\hline
\end{tabular}
\aaf
\end{table}

\todo{Earlier we have shown that the block embedding model with 2 network layers and 20 epochs of training can  
achieve a good performance (Section~\ref{subsec:eval-hyper}), which means 
that it requires five and a half hours (=$(213+275+290+193) \times 20 / 3600$) to train the four models 
on the four training subsets, and each model takes around an hour and a half for training.
With a single network layer, each model only needs about 40 mins for training and can still achieve 
a good performance.}

\subsubsection{Testing Time}

We next investigate the testing time of \bbtool. We are interested in the impacts of the number of 
network layers and the dimension of block embeddings, in particular. {\color{black}Figure}~\ref{tab:TestTime} summaries 
the similarity test on the four testing datasets in Dataset I. The result indicates that the number of 
network layers is the major contributing factor of the computation time. Take the second column as an example. 
For a single-layer LSTM network with the block embedding dimension as 50, it takes 0.41 ms ($=3.530/8722$) 
\zedit{on} average to measure the similarity of two blocks, while a double-layer LSTM network requires 0.76 ms 
($=6.663/8722$) \zedit{on} average.  

\noindent \textbf{Comparison with Symbolic Execution.}
{\color{black}We compare the proposed embedding model 
with one previous basic-block similarity comparison tool which relies on symbolic execution and theorem proving~\cite{luo2014semantics}.} We randomly select 1,000 block pairs and use the symbolic
execution-based tool to measure the detection time for each pair. The result shows that the \bbtool\ runs 
3700x to 140000x faster, and the speedup can be as high as 8000x on average.
 
The reason for the high efficiency of our model is that most computations of \bbtool\ are implemented as 
easy-to-compute matrix operations (e.g., matrix multiplication, matrix summation, and element-wise operations 
over a matrix). Moreover, such operations can be parallelized to utilize multi-core CPUs or GPUs to achieve 
further speedup.

\subsection{Code Component Similarity Comparison}  \label{subsec:bugsearch}

We conduct three case studies to demonstrate how \cctool\ can handle real-world programs 
for \emph{cross-architecture code component similarity detection}.  

\subsubsection{Thttpd}  \label{subsec:cryptofunction}
This experiment evaluated \texttt{thttpd} (v2.25b) and \texttt{sthttpd} (v2.26.4), where 
\texttt{sthttpd} is forked from \texttt{thttpd} for maintenance. Thus, their codebases are 
similar, with many patches and new building systems added to \texttt{sthttpd}. To measure 
false positives, we tested our tool on four independent programs, including \texttt{thttpd} 
(v2.25b), \texttt{atphttpd} (v0.4b), \texttt{boa} (v0.94.13), and \texttt{lighttpd} (v1.4.30). 
We use two architectures (x86 and ARM) and \texttt{clang} with different compiler optimization 
levels (O1-O3) to compile each program. 

We consider a part of the \texttt{httpd\_parse\_request} function as well as the functions
invoked within this code part from \texttt{thttpd} as the query code component, and check 
whether it is reused in \texttt{sthttpd}. Such code part checks for HTTP/1.1 absolute URL and 
is considered as critical. We first identify
the starting blocks both in the query code component and the target program \texttt{sthttpd} 
(Section~\ref{sec:com-comp}), and proceed with the path exploration to calculate the similarity
score, which is 91\%, indicating that \texttt{sthttpd} reuses the query code component. The 
whole process is finished within 2 seconds. However, \texttt{CoP}~\cite{luo2014semantics}  
(it uses symbolic execution and theorem proving to measure the block similarity) takes almost 
one hour to complete. Thus, by adopting techniques in NMT to speed up block comparison, \tool\ 
is more efficient and scalable.

To measure false positives, we test \tool\ against four independently developed programs.
We use the query code component to search for the similar code components in \texttt{atphttpd} (v0.4b), 
\texttt{boa} (v0.94.13), and \texttt{lighttpd} (v1.4.30). Very low similarity scores (below 4\%) are 
reported, correctly indicating that these three programs do not reuse the query code component.

\subsubsection{Cryptographic Function Detection}  \label{subsec:cryptofunction}

We next apply  \tool\ to the cryptographic function detection task. We choose MD5 and AES as the query 
functions, and search for their implementations in 13 target programs ranging from 
small to large real-world software, including \texttt{cryptlib} (v3.4.2), \texttt{OpenSSL} (v1.0.1f), 
\texttt{openssh} (v6.5p1), \texttt{git} (v1.9.0), \texttt{libgcrypt} (v1.6.1), \texttt{truecrypt} (v7.1a), 
\texttt{berkeley DB} (v6.0.30), \texttt{MySQL} (v5.6.17),  \texttt{glibc} (v2.19), \texttt{p7zip} (v9.20.1),
\texttt{cmake} (v2.8.12.2), \texttt{thttpd} (v2.25b), and \texttt{sthttpd} (v2.26.4). We use x86 
and ARM, and \texttt{clang} with O1--O3 optimization levels to compile each program.

\vspace{3pt}
\noindent \textbf{MD5.}
MD5 is a cryptographic hash function that produces a 128-bit hash value. 
We first extract the implementation of MD5 from \texttt{OpenSSL} compiled targeting x86 with 
\texttt{-O2}. 
The part of the MD5 code that 
implements message compressing is selected as the query. 

We use the query code component to search for similar code components from the target programs. 
The results show that \texttt{cryptlib}, \texttt{openssh}, \texttt{libgcrypt}, \texttt{MySQL}, 
\texttt{glibc}, and \texttt{cmake} implement MD5 with similarity scores between 88\% and 93\%.
We have checked the source code and confirmed it.

\vspace{3pt}
\noindent \textbf{AES.}
AES is a 16-byte block cipher and processes input via a substitution-permutation network. 
We extract the implementation of AES from \texttt{OpenSSL} compiled for ARM with  
\texttt{-O2}, and select a part of the AES code that implements transformation 
iterations as the query code component. 

We test the query code component to check whether 
it is reused in the target programs, and found that \texttt{cryptlib}, \texttt{openssh},
\texttt{libgcrypt}, \texttt{truecrypt}, \texttt{berkeley DB},
and \texttt{MySQL} contain AES with the similarity scores between 86\% and 94\%, 
and the others do not. We have checked the source code and obtained consistent results.

\vspace{3pt}
The case studies demonstrate that \cctool\ is an effective and precise tool for cross-architecture 
binary code component similarity detection.

\section{\todo{Discussion}}

\zedit{We chose to modify LLVM to prepare similar/dissimilar basic blocks, as LLVM is well structured
as passes and thus it is easier to add the basic block boundary annotator to LLVM than GCC.}
{\color{black}However, the presented model merely learned from binaries compiled by LLVM. 
We have not evaluated how well our model can be used to analyze binaries in the case binaries are
compiled using diverse compilers.} \zedit{As word embeddings and LSTM are good at extracting
instruction semantics and their dependencies, we believe our approach itself is compiler-agnostic.
We will verify this point in our future work.
}

We evaluated our tool on its tolerability of the syntactic variation introduced 
by different architectures and compiling settings; but we have not evaluated the impact of code 
obfuscation. How to handle obfuscations on the basic block level without
relying on expensive approaches such as symbolic execution is a challenging
and important problem. We plan to explore, with plenty of obfuscated binary basic blocks
in the training dataset, whether the presented model can handle obfuscations by
properly capturing the semantics of binary basic blocks.  \zedit{But it is notable that, 
at the program path level, our system inherits
the powerful capability of handling obfuscations due to, e.g., 
garbage code insertion and opaque predicate insertion, from CoP~\cite{luo2014semantics}.} 

Finally, it is worth pointing out that, as
many prior 
systems are built on basic block comparison or representation~\cite{gao2008binhunt,iBinhunt,luo2014semantics,
pewny2015cross,feng2016scalable}, they can benefit from our block embedding model, which provides 
precise and efficient basic block information extraction and comparison.



\section{Conclusion}

Inspired by Neural Machine Translation, which is able to compare the meanings of sentences 
of different languages, we propose a novel neural network-based basic-block similarity comparison 
tool \bbtool\ by regarding \emph{instructions as words} and \emph{basic block as sentences}.
We thus borrow techniques from NMT: word embeddings are used to represent instructions
and then LSTM is to encode both instruction embeddings and instruction dependencies. 
It is the first tool that achieves both efficiency and accuracy for cross-architecture 
basic-block comparison; plus, it does not rely on any manually selected features. By 
leveraging \bbtool, we propose the first tool \cctool that resolves the cross-architecture
code containment problem. We have 
implemented the system and performed a comprehensive evaluation. 
This research successfully demonstrates that it is promising to approach binary analysis 
from the angle of language processing by adapting methodologies, ideas and techniques in NLP. 

\section*{Acknowledgment}

{\color{black}The  authors  would  like  to  thank  the  anonymous  reviewers
for their constructive comments and feedback.}
This project was supported by NSF CNS-1815144 and NSF CNS-1856380.



%
\balance

\bibliographystyle{IEEEtranS} 
{
\bibliography{code-embedding.bib} 
}

\end{document}